\documentclass[10pt,preprint]{aastex}
\usepackage{amsmath}
\usepackage{graphicx}
    \usepackage{epsfig}
    \usepackage{natbib}
    \usepackage{xcolor}
    \usepackage{arydshln}
    \usepackage{float} 
    \usepackage{changes}

\newcommand{\kms}{{km~s$^{-1}$}}
\newcommand{\ebv}{\ensuremath{E_{B-V}}\ }

\newcommand{\nv}{N\,{\footnotesize V}\ } 
\newcommand{\oi}{O\,{\footnotesize I}\ } 
\newcommand{\feii}{Fe\,{\footnotesize II}\ } 
\newcommand{\hei}{He\,{\footnotesize I}*\ } 
\newcommand{\nai}{Na\,{\footnotesize I}\ } 
\newcommand{\caii}{Ca\,{\footnotesize II}\ } 
\newcommand{\neiii}{[Ne\,{\footnotesize III}]\ } 
\newcommand{\nev}{[Ne\,{\footnotesize V}]\ } 
\newcommand{\mgii}{Mg\,{\footnotesize II}\ } 

\newcommand{\oiii}{[O\,{\footnotesize III}]\ }
\newcommand{\oii}{[O\,{\footnotesize II}]\ }
\newcommand{\halpha}{H\ensuremath{\alpha}\ }
\newcommand{\hbeta}{H\ensuremath{\beta}\ }
\newcommand{\hgamma}{H\ensuremath{\gamma}\ }
\newcommand{\paalpha}{Pa\ensuremath{\alpha}\ }
\newcommand{\pabeta}{Pa\ensuremath{\beta}\ }
\newcommand{\pagamma}{Pa\ensuremath{\gamma}\ }

\newcommand{\nii}{[N\,{\footnotesize II}]\ } 
\newcommand{\sii}{[S\,{\footnotesize II}]\ } 
\newcommand{\ovi}{O\,{\footnotesize VI}\ } 
\newcommand{\civ}{C\,{\footnotesize IV}\ } 
\newcommand{\alii}{Al\,{\footnotesize II}\ } 
\newcommand{\aliii}{Al\,{\footnotesize III}\ } 
\newcommand{\x}[1]{{\color{black} #1}}

\shorttitle{IRAS F11119+3257}
\shortauthors{Pan et al.}

\begin{document}

\title{Discovery of Metastable \hei $\lambda$10830 Mini-broad Absorption Lines and Very Narrow Paschen $\alpha$ Emission Lines in the ULIRG Quasar IRAS F11119+3257}
\author{Xiang Pan\altaffilmark{1}, Hongyan Zhou\altaffilmark{1,2}, Wenjuan Liu\altaffilmark{3}, Bo Liu\altaffilmark{2,1}, Tuo Ji\altaffilmark{1}, Xiheng Shi \altaffilmark{1}, Shaohua Zhang \altaffilmark{1}, Peng Jiang\altaffilmark{1}, Huiyuan Wang \altaffilmark{2}, Lei Hao\altaffilmark{4}}

\altaffiltext{1}{SOA Key Laboratory for Polar Science, Polar Research Institute of China, 451 Jinqiao Road, Shanghai, 200136, China; zhouhongyan@pric.org.cn}
\altaffiltext{2}{Key Laboratory for Researches in Galaxies and Cosmology, Department of Astronomy, University of Sciences and Technology of China, Chinese Academy of Sciences, Hefei, Anhui, 230026, China; mtzhou@ustc.edu.cn, sidpx@mail.ustc.edu.cn}
\altaffiltext{3}{Yunnan Observatories, Chinese Academy of Sciences, Kunming, Yunnan 650011, China
}
\altaffiltext{4}{Shanghai Astronomical Observatory, Chinese Academy of Sciences, Shanghai 200030, China}

\begin{abstract}
IRAS F11119+3257 is a quasar-dominated Ultra-Luminous InfraRed Galaxy, with a partially obscured narrow-line seyfert 1 nucleus. In this paper, we present the NIR spectroscopy of F11119+3257, in which we find unusual Paschen emission lines, and metastable \hei$\lambda$~10830\AA~ absorption associated with the previously reported atomic sodium and molecular OH mini-BAL (Broad Absorption Line) outflow.
Photo-ionization diagnosis confirms previous findings that the outflows are at kilo-parsec scales.
Such large-scale outflows should produce emission lines. We indeed find that high-ionization emission lines (\oiii, \neiii\x{,} and \nev) are dominated by blueshifted components at similar speeds to the mini-BALs. The blueshifted components are also detected in some low-ionization emission lines, such as \oii$\lambda$3727 and some Balmer lines (\halpha, \hbeta, \x{and} \hgamma), even though their cores are dominated by narrow ($FWHM_{\rm NEL} = \x{570\pm40}$\kms) or broad components at the systemic redshift of $z=0.18966\pm0.00006$.
The mass flow rate (\x{230-730}$~M_\odot \rm yr^{-1}$) and the kinetic luminosity ($\dot{E}_k \sim 10^{\x{43.6-44.8}} $erg s$^{-1}$) are then inferred jointly from the blueshifted emission and absorption lines. 
In the NIR spectrum of F11119+3257, we also find that the Paschen emission lines are unique, in which a very narrow ($FWHM=260\pm20~$\kms) component is shown in only \paalpha. \x{This narrow component most probably comes from heavily obscured star formation.} Based on the \paalpha and \pabeta emissions, we obtain an extinction at the $H$ band\x{,} $A_H~>~2.1$ (or a reddenning \x{of} \ebv$~>~$3.7)\x{,} and a star formation rate \x{of} $SFR~>~130\rm M_\odot yr^{-1}$ \x{that resembles} the estimates inferred from \x{the FIR emissions ($SFR_{\rm FIR} = 190\pm90$ $M_\odot$ yr$^{-1}$)}. 
\end{abstract}

\keywords{galaxies: evolution --  galaxies: absorption lines -- galaxies: emission lines -- Galaxies: individual (IRAS F11119+3257)}

\section{INTRODUCTION}
Through decades of studies of local galaxies, the tight correlation between \x{the mass of the} central black holes and \x{the} velocity dispersion of the galaxy bulges (the so-called $M-\sigma$ relation) is established, which indicates close relation between super-massive black holes (SMBHs) and their host galaxies (Ferrarese \& Merritt 2000; Kormendy \& Ho 2013).
On the other hand, observational evidences of quasar feedback have been found (see Fabian 2012 for a detailed review). Both analytical models (Scannapieco \& Oh 2004) and numerical simulations (Sijacki et al. 2007) discovered that quasar feedback is an important ingredient \x{for} regulating galaxy formation, which could lead to the close correlation bewteen SMBHs and their host galaxies. 

As one of the key component of quasar feedback, outflows are found in many quasars. 
Mildly relativistic FeK absorption lines (ultra-fast outflows, or UFOs) are detected in at least $\sim$ 35\% of the radio quiet AGNs, which are regarded as evidences of AGN accretion disk winds (Tombesi et al. 2010).
On the other hand, fast massive OH outflows (maximum speed $\gtrsim$ 1000\kms) are detected in $\sim$ 1/3 of \x{the} nearby galaxy mergers with \textit{Herschel} (Veilleux et al. 2013). 
By comparing the energetics of \x{the} UFOs/(disk winds) and \x{the} massive molecular outflows, Tombesi et al (2015) argues that energy-conserving mechanism is the basis of the quasar outflow feedback.
In addition to the extremely\x{-}ionized FeK absorption line in the X-ray band and \x{the} molecular absorption lines in the FIR, there are copious broad absorption line (BAL) species in the UV-optical spectrum of BAL quasars. Some high\x{-}ionization absorption lines like \ovi, \nv, \civ doublets are nearly universal in BAL quasars, while \aliii, \alii, \mgii and \feii lines are only significant in Low-ionization BALs (LoBALs). \x{The} metastable \hei absorption, on the other hand, are found to be prevalent in \mgii BALs (Liu et al. 2015). And associated atomic broad absorption lines like \nai are also reported (e.g. Liu et al. 2016).
LoBALs generally trace relatively thick or dense outflow clouds, whose physical scale can extend to \x{several kilo-parsecs} (e.g. Chamberlain \& Arav 2015), i.e. could pose strong feedback to the host galaxies.

Sample studies have found that quasar emission lines are generally asymmetric and blueshifted, e.g. in \civ (Richards et al. 2011; Wang et al. 2011) and in \oiii, \nii, \sii\x{,} etc (Komossa et al. 2008; Zhang et al. 2011). \x{On the other hand}, \x{high-ionization BELs in the SDSS quasar composite spectrum (Vanden Berk et al. 2001)} are generally blueshifted as compared to low\x{-}ionization lines. These are usually referred to as signs of emission\x{-}line outflows. \x{In the meantime,} blueshifted emission\x{-}line outflows are found to be associated with LoBALs in some quasars (e.g. Rupke et al. 2013; Liu et al. 2016).
 By analysing blueshifted absorption and emission lines, some properties of the outflow material can be obtained: 1) physical conditions like position, density\x{,} and energetics (Sun et al. 2017)\x{;} 2) geometry (Greene et al. 2012)\x{;} 3) abundances and chemistry (Dunn et al. 2010). These allow us to probe the driving mechanism of quasar outflows and \x{the strength of outflow} feedback to their host galaxies.

IRAS F11119+3257 is a ULIRG ($L_{\rm IR} = 10^{12.58} L_{\odot}$, Kim \& Sanders 1998), and a quasar, or more specifically a partially obscured narrow-line Seyfert 1 (NLS1) galaxy (FWHM$_{\rm \hbeta}$ = 1980 \kms; Komossa et al. 2006) in local universe at $z = 0.18966\pm0.00006$.
It is the key object in Tombesi et al (2015), where both a UFO and a molecular outflow are detected. Thus F11119+3257 is a perfect laboratory for studying physical properties of AGN outflows. In this paper, we report a detailed analysis of the blueshifted ionic/atomic absorption and emission lines in the optical-IR spectrum of F11119+3257\x{. The detection of these lines} helps in constraining \x{physical properties of} the massive molecular outflow. In the meantime, heavily obscured star formation activity in the host galaxy are probed using IR Hydrogen emission lines and PAH/FIR emissions. The connection of the star formation with the galactic\x{-scale} outflow is then discussed. Throughout the paper, all uncertainties are at \x{the} 1-$\sigma$ level if not stated specifically. We adopt \x{the} standard $\Lambda$CDM model with $H_0= 70$~\kms~ Mpc$^{-1}$, $\Omega_{\rm M}=0.3$, and $\Omega_{\Lambda}=0.7$.

\section{OBSERVATION AND DATA REDUCTION}
Photometric and spectroscopic data of F11119+3257 are tabulated in Table~\ref{tbl-1}.
Photometric data from optical (SDSS, York et al. 2000) and IR (the Two Micron All Sky Survey, or 2MASS, Skrutskie et al. 2006; Wide-field Infrared Survey Explorer, or $WISE$, Wright et  al. 2010) bands are collected to study the Spectral Energy Distribution (SED) of F11119+3257.
Multi-epoch photometries are also checked \x{and no significant variability is discovered}. 
Catalina Sky Survey\footnote{http://nesssi.cacr.caltech.edu/DataRelease/} (Drake et al. 2009; ID: CSS\_J111438.9+324133) $V$-band maginitude varied at a level of $\sim$0.08 mag between 2005 May 5 and 2013 Jun 7,
 PTF (Palomar Transient Factory\x{;} Law et al. 2009) magnitude varied at a level of $\sim$0.04 mag between 2013 Jan 13 and 2014 Apr 04, 
 both \x{of the magnitude fluctuations} are consistent with their measurement uncertainties.
And no intra-night varibility in the infrared is found during $WISE$ observations.
The IRAS (The Infrared Astronomical Satellite) photometries at 30$\mu$m, 60$\mu$m, and 100$\mu$m (Moshir \& et al. 1990), and the SCUBA photometry at 850 $\mu$m (Clements et al. 2010) of F11119+3257 are collected to study cold dust emission and star formation activity of the host galaxy.
The radio emission of F11119+3257 is detected by the VLA Faint Images of the
Radio Sky at Twenty-Centimeters Survey (FIRST\x{;} White et al. 1997). With a radio index of $R_{1.4} = 20$, F11119+3257 can be referred to as ``radio intermediate'' (Komossa et al. 2006).

There are a number of archival spectroscopic observations of F11119+3257, 
which are also summarized in Table~\ref{tbl-1}.
In addition, we followed this quasar up with the TripleSpec spectrograph (Wilson et al. 2004)
on the Hale 200-inch telescope (P200) at \x{the} Palomar Observatory on 2013 Feb 24.
Four exposures of 200 seconds are taken in an A-B-B-A dithering mode. 
The 1-D spectrum is then retrieved using the IDL SpexTool package (Cushing et al. 2004).
\x{With} a higher spectral resolution and a broader wavelength coverage, the TripleSpec data thus render higher precision in measuring absorption lines and emission lines than the archival VLT/ISAAC spectrum of F11119+3257. 
All photometric and spectroscopic data are corrected for Galactic extinction at \ebv=0.02 according to the updated dust map of Schlafly \& Finkbeiner et al. (2011).
Before follow-up analysis, all data are deredshifted at $z = 0.18966\pm0.00006$ determined using the narrow emission lines (\S3.2.2), and the laboratory vacuum wavelengths of the lines studied in this paper are adopted from Vanden Berk et al. (2001) (optical), Landt et al. (2008) (NIR), Sturm et al. (2002) (mid-IR [MIR])\x{,} and corresponding references therein.

\section{DATA ANALYSIS}
\subsection{Spectral Energy Distribution and Extinction}
Inspecting the broad-band SED of F11119+3257 from optical to Mid-infrared, we \x{find} its color to be exceedingly red ($(u - W_1)_{\rm F11119+3257} \sim 11.3$) as compared to typical type-1 quasars (i.e. $(u - W_1)_{\rm 3C\ 273} \sim 4.3$) or even typical ULIRGs ($(u - W_1)_{\rm Mrk\ 231} \sim 7.8$). Therefore, heavy extinction in the optical band is expected in  F11119+3257.
Balmer emission line ratio of F11119+3257 also indicates an extinction to the Broad Emission-Line Region (BELR) at $E_{B-V} $ $\approx$ 1.08 (Zheng et al. 2002). F11119+3257 is therefore heavily reddened, and its extinction curve needs to be figured out before modeling of the spectrum. 

\begin{figure}[H]
  \begin{center}
      \includegraphics[width=1\textwidth]{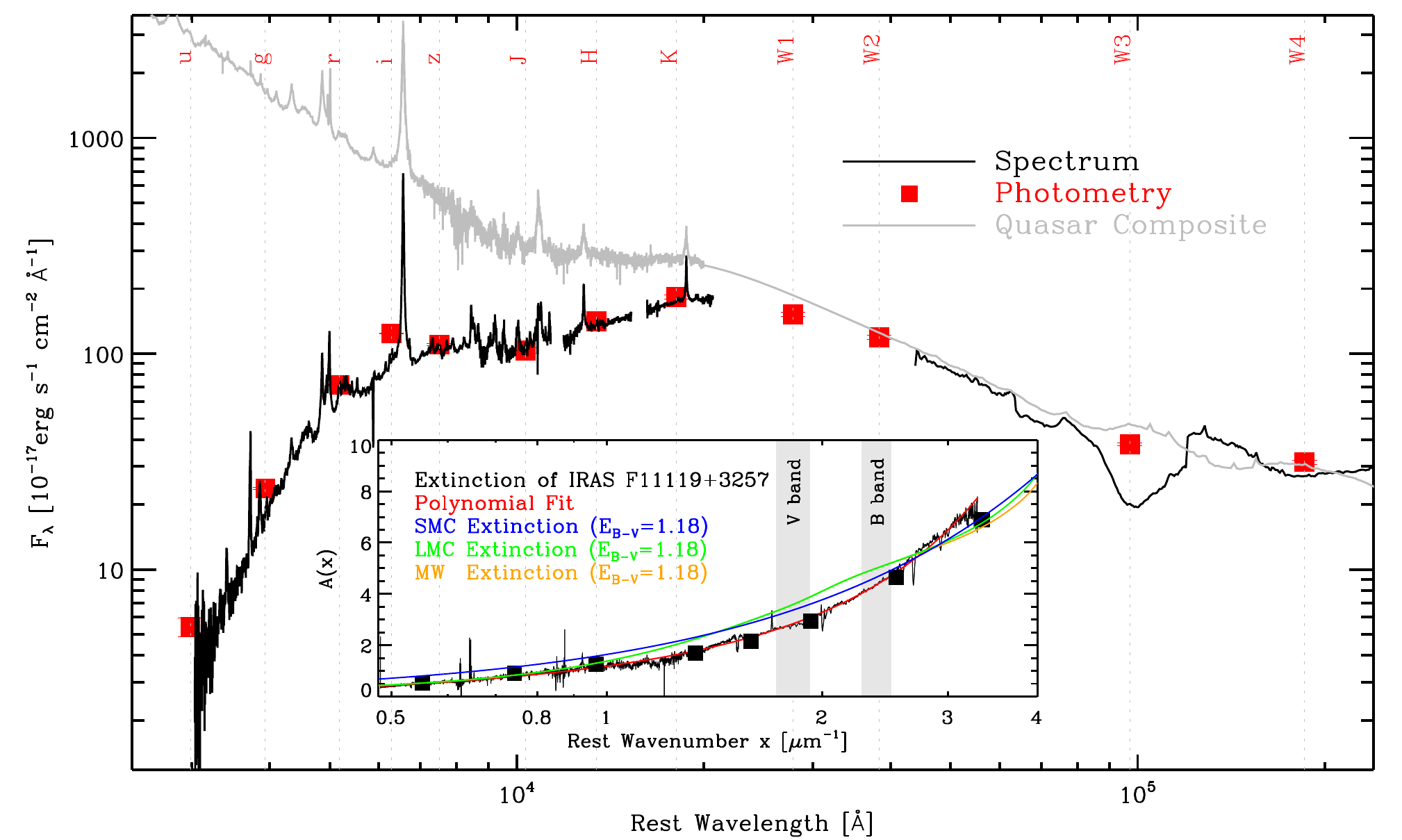}
       \caption{
       SED and Extinction curve (inner panel) of F11119+3257. 
       The spectrum (black lines) and photometric data (red squares) are compared with the quasar 
       composite scaled at $WISE~W_4$ (grey solid line).
       The observed extinction curve of F11119+3257 (black line in the inner pannel) is compared with
       SMC (blue), average LMC (green) and average MW (orange) extinction curves
       at a similar color excess of \ebv=1.18.
      }
  \end{center}
\end{figure}

Since stellar absorption features are not obviously seen in F11119+3257, starlight can be ignored. And scattered AGN radiation, \x{which} usually results in a blue continuum in obscured quasars (Li et al. 2015), is also negligible as the flux density at the blue end of \x{the} spectrum is low. 
It is thus assumed that the optical-IR continuum of F11119+3257 is dominated by the obscured AGN emission.
By comparing the SED of F11119+3257 with the quasar composite (Zhang et al. 2017, a combination of the optical, NIR, and FIR quasar composite from Vanden Berk et al. 2001, Glikman et al. 2006\x{,} and Netzer et al. 2007 respectively), we \x{find} they agree roughly with each other in the MIR ($\sim \lambda \in [3,30]\micron$, see Figure 1). And significant extinction can be seen in both the optical band and the NIR, therefore the physical extent of the obscurer should be large enough \x{to} fully cover both disk continuum and hot/warm blackbody radiation from the torus.
Since the optical-IR SED of F11119+3257 is dominated by \x{the} transmitted AGN radiation, we follow the conventional method of assuming the quasar composite as \x{the} extinction\x{-}free spectrum in deriving the extinction curve. After scaling the \x{quasar} composite to F11119+3257 at $WISE\ W_4$, its extinction curve is then calculated, $A_\lambda = -2.5\times log (F_{\lambda}^{\rm F11119+3257} / F_{\lambda}^{\rm composite})$ (inner panel of Figure 1). This extinction curve is then fitted with a cubic polynomial, which is
$A(x) = -0.22 + 1.22 x +0.12 x^2 + 0.07x^3$, where ${\rm}\ x= (\lambda/1\mu m)^{-1} $. 
Consisting of only 4 polynomial coefficients, the model of extinction curve is \x{so} simple \x{that the effect of} high\x{-}frequency spectral noises, i.e. sharp dips and peaks around emission and absorption lines\x{, is largely avoided}.
After applying the response curve of $B$, $V$ filter (Johnson \& Morgan 1953), broad\x{-}band extinction and color excess of F11119+3257 are calculated: $A_V = 2.78$, \ebv = 1.18\x{. The extinction value is similar to that estimated for the BELR of} $E_{B-V} $ $\approx$ 1.08 (Zheng et al. 2002).
Study\x{ing} of the First-2MASS red quasars shows that the $E_{B-V}$ estimated for emission\x{-}line regions are relatively more uncertain as compared to \x{the} $E_{B-V}$ estimated for AGN continuum (Glikman et al. 2007). The reason is that properties of quasar BELs are diverse and could deviate from Case\x{-}B recombination in certain objects. On the other hand, Dong et al. (2008) found that the broad-line \halpha/\hbeta line ratio of blue AGNs peaks at 3.06 (with a relatively small spread of 0.22), suggesting that Balmer decrement can indicate \ebv in normal AGNs. Thus the remarkable agreement between the extinction evaluated for both the continuum and the BELR indicates that F11119+3257 follows general properties of AGNs and the extinction evaluations are reasonable.
In comparison with extinction curves of SMC, average LMC\x{,} and average Milky Way at \ebv=1.18 (Prevot et al.1984\x{;} Fitzpatrick 1986\x{;} Fitzpatrick \& Massa 2007), we find that the extinction curve of F11119+3257 is relatively steeper. Such kind of steep extinction curve is similar to those reported in many other BAL quasars (e.g. Jiang et al. 2013; Zafar et al. 2015).

\subsection{Emission Lines}
\subsubsection{Continuum and \feii Multiplets}

Before modeling Emission Lines, broad spectral features, i.e. the continuum and the \feii multiplets\x{,} needs to be subtracted. For the featureless continuum, we first select multiple continuum regions with widths of $\Delta\lambda/\lambda = 0.001$
(triangles in the top panel of Figure 2). We then interpolate these continuum points into the entire spectrum in \x{log $\lambda$-log $\rm F_\lambda$ } space (dashed lines in Figure 2.1) and obtain the \x{baseline model of the} underlying continuum.

The \x{baseline} underlying continuum is then subtracted, and the \feii multiplets are fitted in the four spectral regions showing prominent \feii emissions. They are [4400,4700]\AA, [5120,5700]\AA, [6000,6270]\AA\ and [6950,7350]\AA\ in quasar's rest frame (grey shaded regions in Figure~2.1). 
Following Dong et al. (2011), a two-component analytic form of \feii emission lines incorporating \x{the} VJV04 \feii template (V{\'e}ron-Cetty et al. 2004) are used as the model.
The narrow one of the two \feii components is assigned with a \x{G}aussian profile.
\x{Both Gaussian and Lorentzian profiles are tested for the broad \feii component. Similar to many other NLS1s, the Lorentzian profile approximates the data better.}
On the other hand, though, it is found that this \feii model \x{cannot} fit the data well, as the \feii flux ratio in the \feii windows of F11119+3257 is quite different to that of the VJV04 template:
after the removal of the interpolated continuum, the integrated \feii flux ratio of F11119+3257 at the  $\lambda$4570 ([4400,4700]\AA) and the $\lambda$5250 ([5120,5700]\AA) windows is $\rm F_{4570}/F_{5250} = 0.35$, much lower than \x{that of} the template quasar I Zw 1, $\rm F_{4570}/F_{5250} (I Zw 1) = 0.92$. The relatively low flux of \feii emission at shorter wavelengths in F11119+3257 indicates significant reddening of the \feii emission. We thus apply the cubic polynomial extinction curve of F11119+3257's continuum (\S3.1) to the \feii multiplets, with a scaling factor variable $E_{B-V}^{\rm \feii}$.
Even though continuum windows are selected to avoid \feii lines, some continuum points can still be contaminated by \feii emissions. In order to solve the entanglement between continuum and \feii multiplets, we performed the modeling iteratively:\\
Step 1, interpolate continuum points (orange triangles in Figure~2.1) \x{to the spectrum in log $\lambda$-log $\rm F_\lambda$} space to obtain continuum model (orange dashed lines in Figure~2.1).\\
Step 2, subtract the continuum model and fit the \feii emission in the four \feii windows (orange solid line in Figure~2.1).\\
Step 3, subtract the \feii emission model at continuum points and obtain the corrected continuum fluxes (red triangles in Figure~2.1).\\
Step 4, repeat Step 1-3 until the fitted \feii fluxes converge (\x{with a precision of 1\%}).\\
Finally, at a best-fit \x{of} $\rm \chi^2/DOF\sim 1.1$, modeled continuum and \feii emissions are obtained  (red lines in Figure~2.2). The best-fit profile parameters of the \feii emission lines are listed in Table~2, and extinction to the \feii emission is calculated to be $E_{B-V}^{\rm \feii}$ = $1.4\pm0.1$.
The uncertainty of continuum+\feii multiplets is shown with pink shaded areas in the lower three panels of Figure~2, which is derived with a bootstrap approach. The continuum+\feii multiplets model is then subtracted before follow-up analysis of \x{the} emission lines.

\begin{figure}[H]
  \begin{center}
      \includegraphics[width=1\textwidth]{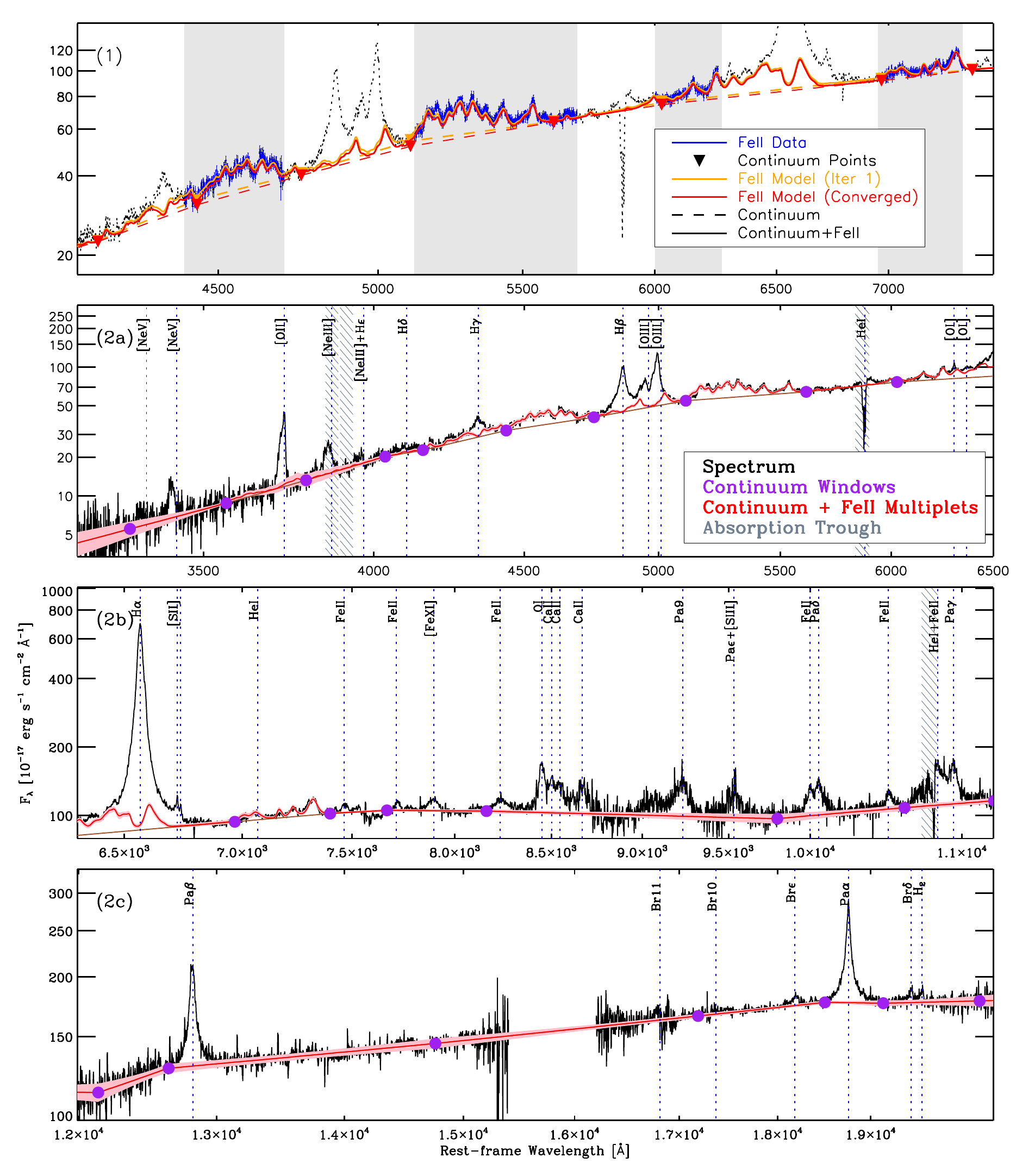}
       \caption{
       Modeling of the continuum and the \feii multiplets are shown in panel 
       (1). Data points in the four 
       \feii windows are colored blue. The results w/o been modeled iteratively
        are colored red and orange respectively (see \S3.2.1 for a detailed description).
       Modeled continuum and continuum+\feii are presented with dashed and solid lines.
       In panel (2a), (2b), (2c), the modeled continuum points are indicated with purple dots.
       The best-fit and the uncertainty of continuum + \feii multiplets model
       are shown \x{with} red lines and pink shades.
       We mark laboratory wavelengths of common emission lines with vertical lines. 
       Spectral regions shaded by tilted lines indicate spectral regions affected by 
       Broad Absorption Lines (BALs, \S3.3). The marked BALs are
       \hei$\lambda$3889, \caii$\lambda$3934, \nai$\lambda\lambda$5891,5897 in panel (2a),
       and \hei$\lambda$10830 in panel (2b).
      }
  \end{center}
\end{figure}

\subsubsection{Broad Emission Lines and Narrow Emission Lines}

In the bottom three panels of Figure~2, an overview of F11119+3257's emission lines is shown. We mark the laboratory wavelengths of the emission lines at a rest frame of $z=0.18966\pm0.00006$ (see detailed description below) with dotted vertical lines. 

At the first glance, the peak of strong Balmer, Paschen, and Brackett emission lines matches the dotted vertical lines. Also, the peaks of low-ionization forbidden lines like \oii $\lambda$ 3727 and \sii $\lambda\lambda$6718,6732 match their dotted line marks, indicating that they share a similar systemic redshift.
Since no obvious stellar absorption is detected, low-ionization Narrow Emission Lines (NELs) appear to be the best redshift indicator of F11119+3257.
\oii $\lambda$ 3727 emission line is commonly used as a redshift indicator. However, this doublet in F11119+3257 is highly asymmetric (Figure 5.4), indicating that significant portion of it is contributed by non-NEL components. Thus the systemic redshift of the quasar cannot be measured directly with the \oii $\lambda$ 3727 emission line.
Other optical low-ionization forbidden emission lines, i.e. \oi $\lambda$ 6302, \nii $\lambda\lambda$6549,6585 and \sii$\lambda\lambda$6718,6732 are blended with the prominent \halpha emission line.
We thus have to model these narrow emission lines together with \halpha in the observed wavelength range of [7471,8090]\AA ([6280,6800]\AA\ at $z=0.18966$).
During the modeling, all these narrow emission lines are assumed to be Gaussians with the same $FWHM$, and their central wavelengths ratios are fixed as their laboratory wavelengths ratios.
The flux ratio of the \nii doublet is fixed at $F_{\rm \nii\lambda6549}/F_{\rm \nii\lambda6585}=1/2.96$.
On the other hand, five Gaussians (one narrow component at $FWHM<1000$\kms~ and four broad components at $FWHM>1000$\kms) are assumed for \halpha to ensure a reasonable fit and a reliable 
determination of the relatively weak NELs.
Redshift of the NEL system is found to be $z=0.18966\pm0.00006$\x{,} which is regarded as the systemic redshift of F11119+3257. The modeled Gaussian width is $FWHM=570\pm40$\kms (Table~2).
Fitting results are shown in panel (1a) of Figure~3. And modeled weak NELs are presented together with the \feii multiplets model and \x{the} residual in panel (1b).

As shown in Figure~2, the most prominent broad Hydrogen emission lines are \paalpha, \pabeta, \halpha, and \hbeta. Among them, \halpha is blended with several weak NELs, and \hbeta is blended with the blueshift \oiii $\lambda\lambda$ 4959,5007 emission lines.
After removal of the the modeled NELs, line profile of \halpha is obtained.
Line spectrum around \hbeta and \oiii \x{is} shown in panel (2a) of Figure~3. And the removed \feii multiplets model is presented in panel (2b). The subtraction of the \feii multiplets seems reasonable and the line profile\x{s} of \hbeta and \oiii can be reliably obtained. For the \hbeta+\oiii blend, we begin with solving the line profile of \x{the} \oiii doublet.
After \x{scaling} the \halpha profile to match the peak of \hbeta emission (panel 2a of Figure~3), we found that in the rest-frame wavelength range of [4940,5020] (grey shaded region), the contribution of \hbeta emission is negligible (weaker than \x{the spectral flux fluctuations}). Therefore, the emission of \oiii $\lambda\lambda$ 4959,5007  should dominate in this spectral region. In addition, the bulk flux of \oiii $\lambda$ 5007 lies in this \x{spectral} region, \x{as well as} the peak of \oiii $\lambda$ 4959. By convention, the line profile of \oiii $\lambda$ 4959  and \oiii $\lambda$ 5007  are assumed to be the same and the line ratio are fixed at $F_{\rm \oiii\lambda4959 }/F_{\rm \oiii\lambda5007 }$=1/2.98 (Storey \& Zeippen 2000). We then solve the the line spectrum in this region pixel by pixel, and obtain the emission line profile of \oiii $\lambda$ 5007  and \oiii $\lambda$ 4959  (blue and green solid line in panel 2a of Figure~3). Finally, after the subtraction of \oiii $\lambda\lambda$ 4959,5007, the line profile of \hbeta is derived.

\begin{figure}[H]
  \begin{center}
      \includegraphics[width=0.5\textwidth]{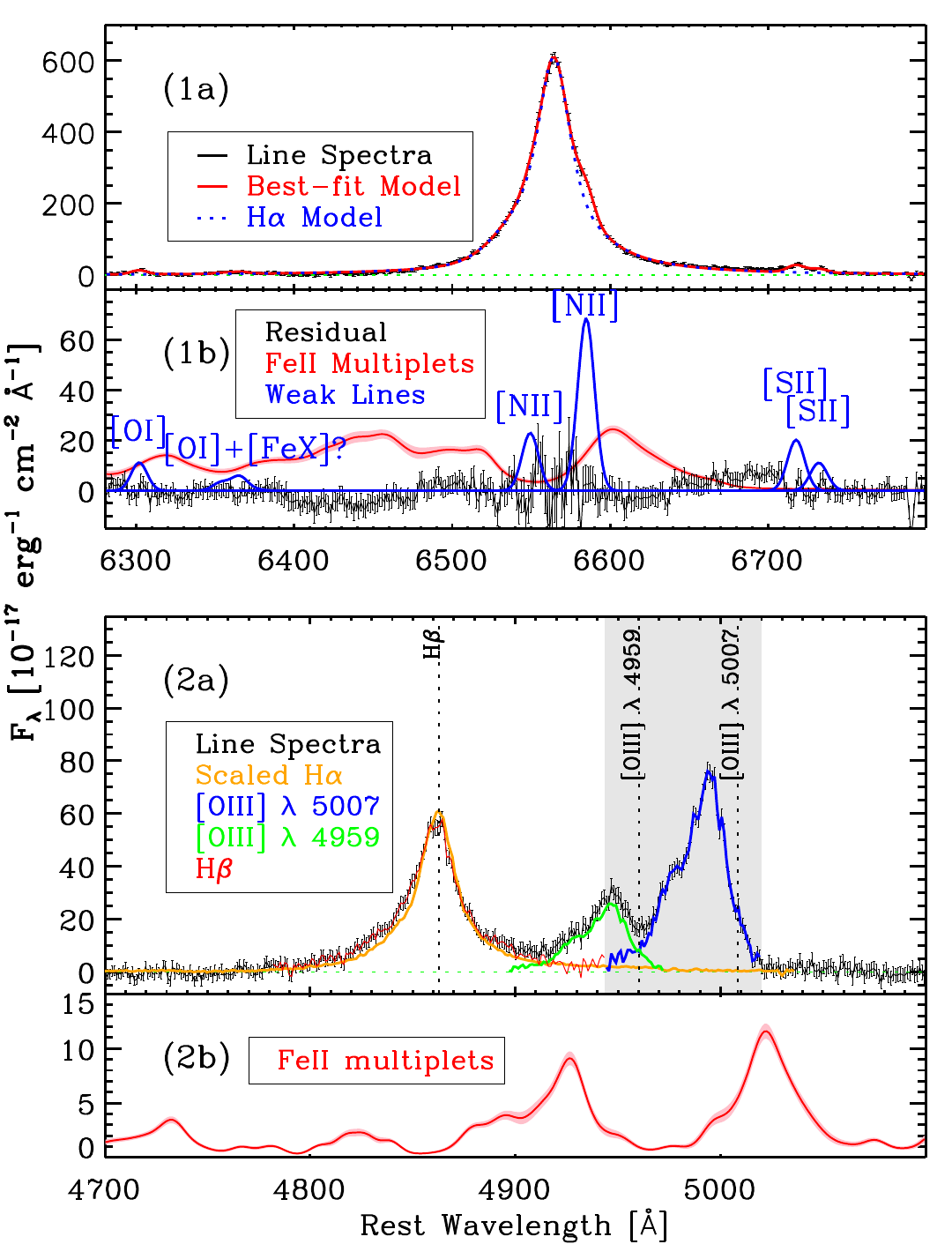}
       \caption{
       Modeling of the \halpha blend is shown in panel (1a). 
       Although the modeled NELs are weak, they can still be distinguished from \feii 
       multiplets and model residuals as shown in pane (1b).
       The \hbeta and \oiii blend is shown in panel (2a). 
       The profile of \oiii $\lambda$ 5007  (blue solid line) 
       is derived in the grey shaded region,
       where \feii emission (panel 2b) is reliably removed and contribution from \hbeta is 
       negligible. The derived \oiii $\lambda\lambda$ 4959,5007  emission is subtracted and
       the \hbeta emission line (red solid line in panel 2a) is obtained.
      }
  \end{center}
\end{figure}

The flux-weighted centroid velocities and $FWHM$s of these four Hydrogen emission lines are measured and shown in Table~2, which differ significantly from each other.
As the wavelength decreases from \paalpha through \pabeta and \halpha to \hbeta, the line centroids vary gradually from at $\sim$0 \kms~ to be blueshifted at $\sim$
100 \kms~ in quasars rest frame . And the $FWHM$ increases gradually from $\sim$900 \kms\  to $\sim$1450 \kms. Such a trend of their profile changing suggests that these Hydrogen emission lines are not dominated by a same single component.
Being permitted lines of the most abundant element in the universe, Hydrogen Balmer/Pashen emissions can be generated in various astrophysical environments. Thus, the emission line systems in F11119+3257: the Narrow Emission Lines, the Broad Emission Lines (BELs), and the blueshifted Outflow Emission Line (OEL, \S3.2.3) systems may all contribute to these observed Hydrogen emission. \x{And different intensity ratios of these emission-line systems can lead to the line-profile variations observed among these Hydrogen lines.}
A brief comparison among the Hydrogen lines is then made as an attempt to investigate the emission\x{-line} components.

Despite their differences, the red wings of these Hydrogen lines resemble each other (in a velocity range of [200,4000]\kms~ in quasar's restframe). This is reasonable since the line flux in this velocity region is dominated by only the BEL system, which leads to the similarity of line profiles in this region. 
In Figure~4, different Hydrogen lines are scaled by their fluxes in their red wings (in a velocity range of [200,4000]\kms), and their profiles are compared. 
The \paalpha emission line has the largest wavelength \x{and thus} lowest extinction, i.e. least affected by obscuring dust.
We therefore use the line profile of \paalpha as a reference since it preserves best the BEL component among these Hydrogen lines. The line profile of \paalpha and \pabeta are similar in both their blue and red wings, while a narrow peak is found in only \paalpha and none of the other Hydrogen lines.
The difference between \paalpha \x{and} Balmer lines lies in two aspects (Figure~4.2, 4.3, 4.4). 1) The sharp and narrow peak of \paalpha appears prominent if \paalpha is compared with Balmer lines. 2) Significant residual fluxes are found in the blue wings of Balmer lines if scaled \paalpha profile is subtracted.
Using \paalpha as a baseline model for the broad and narrow Hydrogen emission line component, the residual fluxes in the blue wings of Balmer lines are very likely associated with \x{the} other blueshifted emission lines like \oiii, \neiii, and \nev.
Since the Balmer BELs are suppressed more severely by dust extinction than the Pashen BELs, the relatively weak blueshifted Emission line components can be revealed in the blue wings of the Balmer lines.
If this is the case, the OEL component then should be more easily observed in lines whose BEL endures stronger extinction, i.e. at shorter wavelengths, as in the case of the obscured quasar OI 287 (Li et al. 2015).
This phenomenon is indeed observed, the \x{relatively strength of residual flux in blue wings of \hbeta is noticeably stronger than \halpha.} 
And for \hgamma at a even shorter wavelength, the residual in the blue wing can be clearly seen under low signal-to-noise ratio even if the potential \oiii $\lambda$ 4363  emission is treated as part of \hgamma emission in the red wing.
Since this residual flux is most clearly observed in \hbeta, this blueshifted component is retrieved by subtracting the scaled \paalpha line profile, and will be investigated later in the follow-up section.

On the other hand, we find that the centers of all Hydrogen emission lines are flat except that a weak but prominent narrow emission is found in the peak of \paalpha. By subtracting the profile of \halpha (the Hydrogen line with the highest signal-to-noise ratio), the narrow \paalpha peak is detected (brown line in Figure 4.2) at a flux of $F_{\rm \paalpha peak} = 370\pm90 \times 10^{-17} \rm erg s^{-1} cm^{-2}$. And the width of the peculiar narrow emission component is  ($FWHM_{\rm \paalpha peak} = 260\pm$20~\kms), significantly lower than the NEL component ($FWHM_{\rm NEL} = 570\pm$40~\kms). We then check this narrow Paschen emission in \pabeta. By subtracting the scaled \halpha from \pabeta, the excess flux at the peak of \pabeta is found to be $F_{\rm \pabeta peak} = 40\pm60 \times 10^{-17} \rm erg s^{-1} cm^{-2}$, consistent with non-detection. Therefore, the line ratio \pabeta/\paalpha for this narrow emission is quite low ($F_{\rm \pabeta peak}/F_{\rm \paalpha peak} < 0.15$) if case-B flux ratios of ($F_{\rm \paalpha}/F_{\rm \pabeta} \approx 0.5$) is considered (Hummer \& Storey 1987). Positively detected only in \paalpha, this anomalous narrow Paschen emission is thus much weaker at shorter wavelength as compared to model predictions. The most plausible explanation to this phenomenon is that this Paschen emission-line component is heavily reddened. By assuming an extinction law of either MW (Fitzpatrick \& Massa 2007) or SMC (Prevot et al.1984), an $H$ band ($\lambda_{\rm eff} \approx 16620\AA$) extinction of $A_H > 2.1$ (corresponding to \ebv $~>~3.7$) can be figured. The low velocity dispersion and high extinction makes this emission line component distinct from all the other emission line systems verified in the quasar F11119+3257, and it will be discussed later in \S4.3.


\begin{figure}[H]
  \begin{center}
      \includegraphics[width=0.5\textwidth]{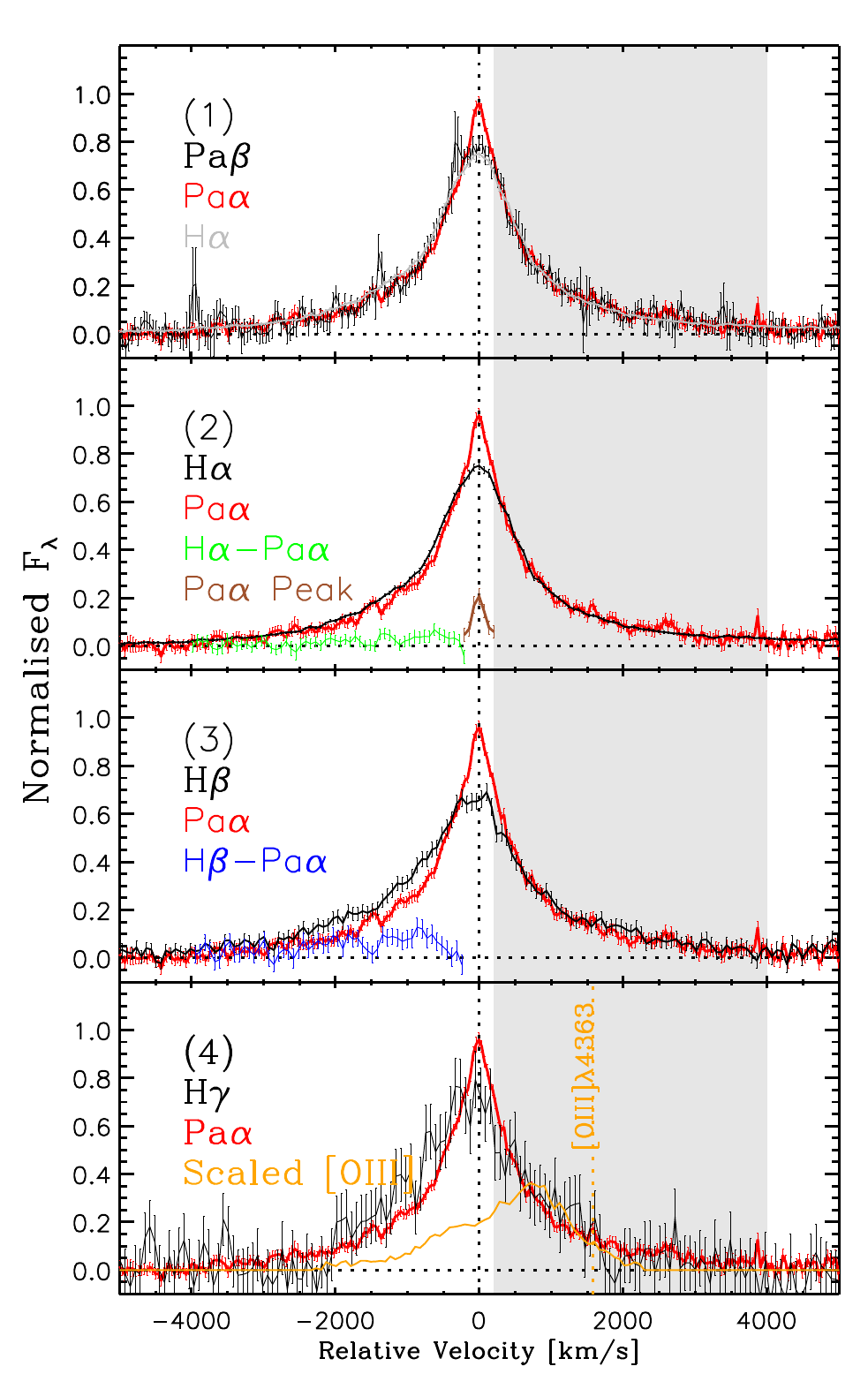}
       \caption{
       From top to bottom panels, emission line profiles of \pabeta, \halpha, \hbeta, and
       \hgamma are compared with that of \paalpha. 
       All emission lines are normalized by their integrated
       flux in the grey shaded region ([200,4000]\kms~ in velocity space).
       A sharp narrow peak unique in \paalpha is found (brown lines), and significant residual fluxes in 
       the blue wings of \halpha (green lines), \hbeta (blue lines), and \hgamma are detected.
      }
  \end{center}
\end{figure}

\subsubsection{Outflow Emission Lines}
As shown in the lower panels of Figure~2, the high-ionization forbidden lines like \oiii $\lambda\lambda$4959,5007  and \nev $\lambda$3427  are all significantly blueshifted in the rest frame of the quasar. Significant flux excess in the blue wings are also detected in \oii $\lambda$3727, \neiii $\lambda$ 3869,  and \neiii $\lambda$ 3968. On the other hand, blueshifted residual fluxes in \halpha and \hbeta (\S3.2.2) are found. They can all originate from the outflow gas.

The blushifted emission lines, \oiii $\lambda$5007, \oii$\lambda$3727, \neiii $\lambda$3869, \neiii$\lambda$3968, \nev$\lambda$3427, as well as the residual \halpha and \hbeta emission are plotted together in the Figure~5.
It is found that for all these lines, the majority of blueshifted emission are in a similar velocity range of $~\sim$[-3500,0]~\kms.
For \oiii $\lambda$5007, \oii$\lambda$3727, and \neiii$\lambda$3869, significant amount of flux at v=0~\kms~ are detected, which can come from the NEL fluxes centered at $v$ = 0 \kms (\S3.2.2). 
This should be removed before analyzing the blueshifted emission line system.
We assume that the NEL component of \oiii $\lambda$5007  and \oii$\lambda$3727  to be symmetric about v=0~\kms, and the flux at $v >$ 0 \kms~ are dominated by the NEL component. The ``NEL'' component (green dashed line in Figure 4) of \oiii$\lambda$5007  and \oii$\lambda$3727  are then obtained and subtracted, which leaves the blueshifted emission line components (blue dashed lines in Figure 5).

\begin{figure}[H]
  \begin{center}
      \includegraphics[width=0.5\textwidth]{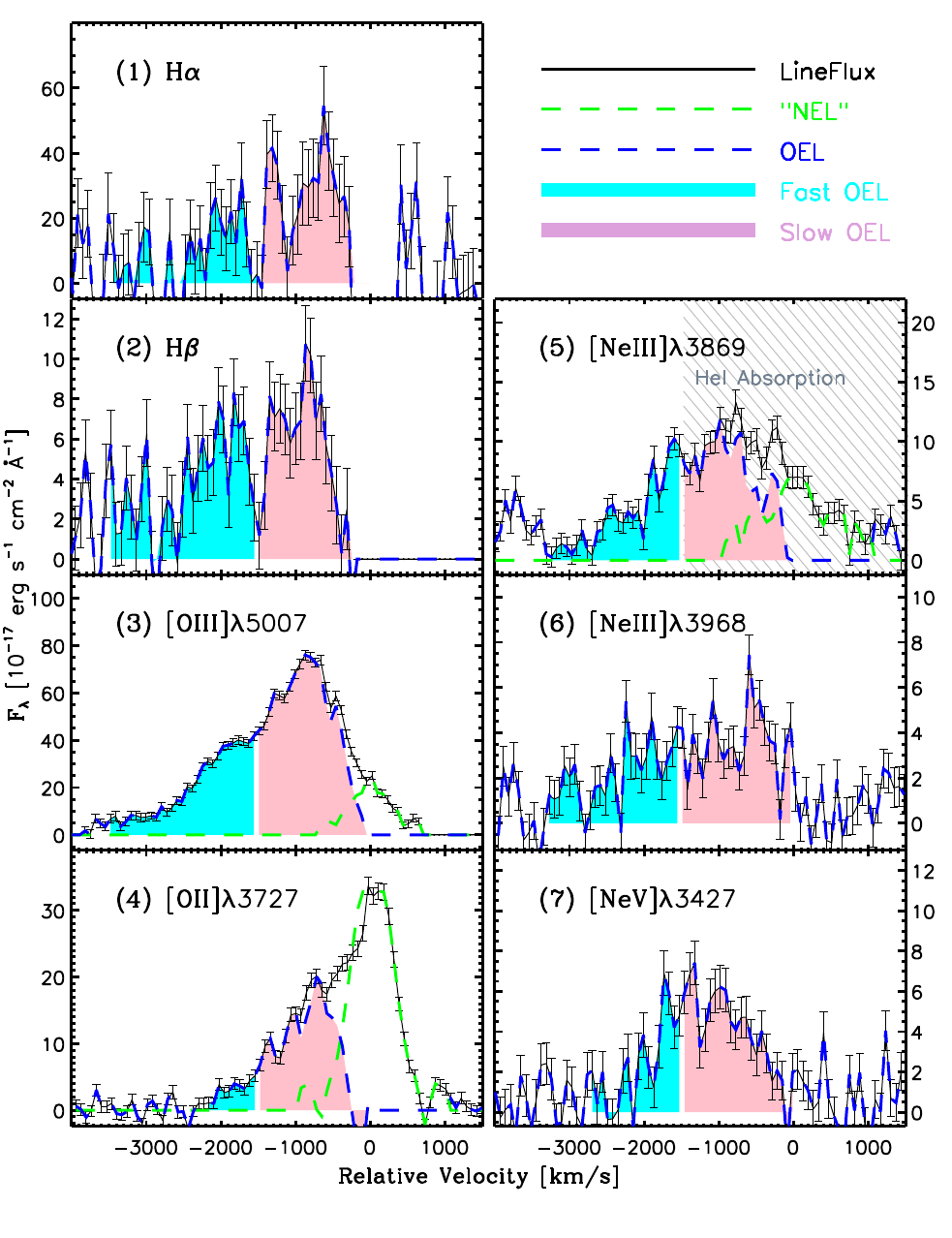}
       \caption{
       Blushifted emission lines in F11119+3257, i.e. (1) residual \halpha, (2)residual \hbeta, 
       (3) \oiii $\lambda$5007, (4) \oii$\lambda$3727, (5) \neiii$\lambda$3869 
       (6) \neiii$\lambda$3968, (7) \nev$\lambda$3427  
       are shown in the figure. The symmetric ``NEL'' system centered 
       at $v =$ 0 \kms~ is presented with green dashed line. The fast and slow OEL components
       are shaded cyan and pink respectively, and their total fluxes are marked with
       blue dashed lines. The slow OEL component of \neiii$\lambda$3869  can be affected
       by \hei $\lambda$ 3889  mini-BAL (shaded with grey lines).
      }
  \end{center}
\end{figure}

This blueshifted emission line system, or the so-called \oiii ``blue outliers'', are usually referred to as tracers of outflows (Komossa et al. 2008). We thus call them Outflow Emission Lines (OELs) in the follow-up analysis.
This OEL system is broad ($FWHM_{\rm \oiii \lambda 5007 } = 1514$~ \kms). In \oiii$\lambda$5007, the OEL with the highest spectral quality, a peak at $v \sim$ -800 \kms~ is found. In addition to this low-speed peak, another broad component can be found that is centered at around $v \sim$  -2000 \kms. It seems that the \hbeta and \neiii OEL also follow such a two-component scenario. We thus divide the OEL system into a high-speed component (fast OEL; $v \in$ [-3500,-1500] \kms) and a low-speed component (slow OEL; $v \in$ (-1500,0] \kms). The integrated fluxes of both the fast OEL and the slow OEL components, and flux-weighted average velocities for each lines are listed in Table 3. Notice that the slow OEL component of \neiii $\lambda$3869  can be contaminated by the \hei $\lambda$ 3889  mini-BAL (marked region in panel 5 of Figure 5), the integrated flux in this region is therefore a lower limit estimation.

\subsection{Broad Absorption Lines and Column Densities}

\begin{figure}[H]
  \begin{center}
      \includegraphics[width=0.52\textwidth]{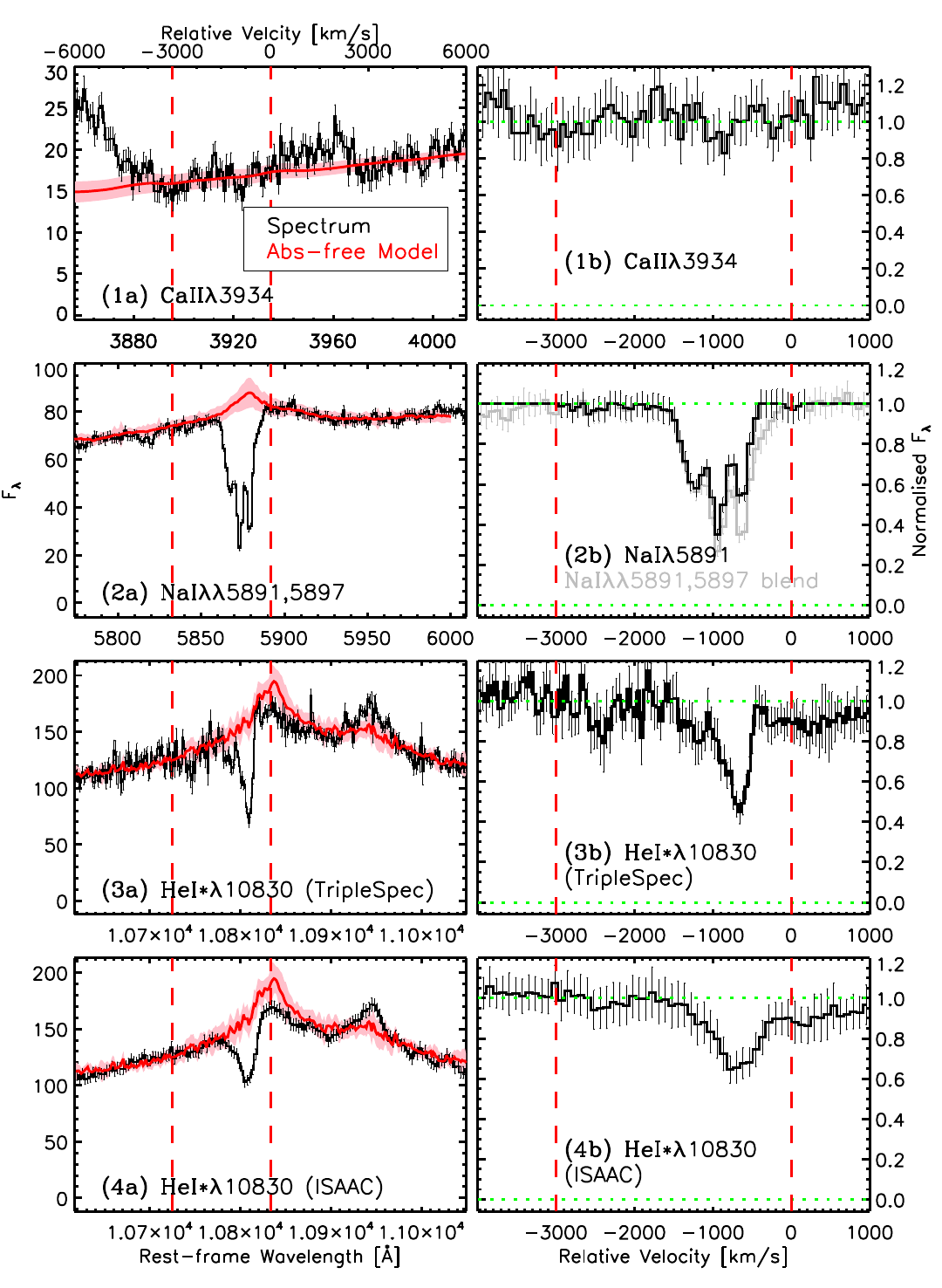}
      \caption{
       Spectrum of F11119+3257 around the \caii$\lambda$~3943,
       \nai$\lambda\lambda$ 5891,5897  and \hei$\lambda$~10830  min-BALs as well as absorption-free
       models (red lines) and model uncertainties (pink shades) are shown in the left panels.
       Normalized spectrum are shown in the right panels. By integrating AOD in a velocity 
       range of [-3000,0]~\kms (red vertical dashed lines), 
       column densities are derived: log $N_{\rm Ca^+}/\rm cm^{-2} < 14.0$  
       log $N_{\rm Na^0}/\rm cm^{-2} = 13.7 \pm 0.1 $ 
       and  log $N_{\rm He^0(2^3S)}/\rm cm^{-2} > 13.2$.
       }
  \end{center}
\end{figure}
Besides the \nai $\lambda\lambda$ 5891,5897  mini-BAL (Rupke et al. 2005) and the blueshifted molecular OH absorption line (Veilleux et al. 2013), we noticed prominent metastable \hei $\lambda$ 10830  and weak \caii $\lambda$3934  absorption lines in the optic-NIR spectrum of F11119+3257. 
The absorption troughs lie in a velocity range of $v\sim[-3000,0]$ \kms (centered at $\sim -1000~$\kms) relative to the systemic redshift of $z = 0.18966$, very similar to the OELs (Table-3). As presented in \S3.1, the optical-NIR spectrum of F11119+3257 are dominated by the transmitted AGN radiation. The dust absorber appears to fully cover the AGN radiation in our sightline (covering factor is about 1), since both the reflected AGN emission and the starlight from host galaxy are negligible. 

For the weak absorption of \caii $\lambda$ 3934, no prominent emission lines are found around the absorption trough. The continuum + \feii model described in \S3.2.1 is a reasonable normalization in this spectral region (Figure~6.1a). Based on the normalized \caii $\lambda$ 3934  absorption profile (Figure~6.1b), we integrate the apparent optical depth (AOD, Savage \& Sembach 1991) in the velocity range of $v \in~$[-3000,0]~\kms~ and derive the upper limit of the ion Ca$^+$'s column density to be log $N_{\rm Ca^+}/\rm cm^{-2}<$14.0.
On the other hand, the \nai $\lambda\lambda$5891,5897  blend and the \hei$\lambda$10830  absorption are close to emission lines (\ion{He}{1} $\lambda$5877  and \hei$\lambda$10830 +\pagamma respectively), which makes simple continuum model invalid in approximating the absorption-free spectrum. Following Liu et al. (2015), we apply pair-matching method to normalize these two absorption lines. For the \nai $\lambda\lambda$5891,5897  blend, spectra of SDSS DR12 quasars (P{\^a}ris et al.2017) at $z <$ 0.5 are used as templates. 
Similar to Liu et al.(2016), we fix an optical depth ratio of $\tau_{\rm \nai \lambda5891 }/\tau_{\rm \nai \lambda5897 } = 2$ and obtain the absorption line profile of \nai $\lambda$5891  (black line in Figure~6.2b). By integrating AOD, the column density of atomic sodium is found to be log $N_{\rm Na^0}/\rm cm^{-2} = 13.7 \pm 0.1$.
Pair-matching method is then preformed to normalize the \hei$\lambda$10830  min-BAL, with 92 infrared quasar spectra from infrared quasar spectral atlases (Glikman et al. 2006, Riffel et al. 2006, Landt et al. 2008). We assume a covering factor of 1 and measure the column density of atomic helium in the metastable 2$^3S$ state via calculating AOD. The results based on the P200/Triplespec and the VLT/ISAAC data are log $N_{\rm He^(2^3S)}/\rm cm^{-2} (TripleSpec) = 13.4\pm0.2$ and log $N_{\rm He^0(2^3S)}/\rm cm^{-2} (ISAAC) = 13.3\pm0.3$, consistent with each other. Provided that the Triplespec data has higher resolution and shows more detailed absorption features, we adopt the more reliable Triplespec result in the follow-up analysis.
Since emission line spectrum around \hei$\lambda$10830  is complex, and the number of template spectra is limited, the normalization is not as good.
In addition to the \hei$\lambda$~10830 +\pagamma BEL, there can be significant \hei$\lambda$10830  OEL (in the velocity range of $v\in[-3500,0]$~\kms described in \S3.2.3) in the absorption trough.
In the meantime, part of the IR radiation comes from blackbody components (\S3.1), which may not be fully covered by the absorbing gas. In view of these facts, the directly normalized \hei$\lambda$10830  absorption can be significantly underestimated, it is thus regarded as the lower limit of the column density for the metastable atomic helium, i.e. log $N_{\rm He^0(2^3S)}/\rm cm^{-2} > $13.2. In view of the uncertainty in measuring \hei $\lambda$10830  absorption, the corresponding \hei $\lambda$3889  absorption remains unknown. The potential \hei $\lambda$3889 absorption trough lies just beneath the \neiii $\lambda$ 3869 slow OEL component. Thus the integrated flux of \neiii $\lambda$ 3869  slow OEL can only be treated as a lower-limit estimate (Figure 5.5).

\subsection{FIR Continuum and Emission Lines}

The Spitzer IRS spectrum of F11119+3257 obtained from ``Cornell Atlas of Spitzer/IRS Sources'' (CASSIS LR7; Lebouteiller et al. 2011) is shown in Figure~7. Significant silicate absorption at $\sim~$9.7~\micron\ is found, as well as several conspicuous PAH emission features.
We modeled the Spitzer spectrum of F11119+3257 with PAHFIT (v1.2). PAHFIT is an IDL tool that decomposes Mid-Infrared spectrum into reddened starlight, thermal dust continuum, and various emission lines. More detailed setup of the modeling can be found in Smith et al. (2007).
Based on the best-fit result of PAHFIT, the combined luminosity of PAH 6.2 \micron\ and PAH 11.2 \micron\ is $L_{\rm PAH (6.2 \micron+11.2 \micron)}$ = 5.0$\pm0.6\times 10^{43}$ erg s$^{-1}$.
Applying the relation of $SFR$($M_\odot$, yr$^{-1}$)=1.18$\times$10$^{−41}$ $L_{\rm PAH (6.2 \micron+11.2 \micron)}$ (typical error is of the order ∼50\%; Farrah et al. 2007), the star formation rate ($SFR$) of F11119+3257 is estimated to be $SFR_{\rm PAH}$=600$\pm$300 $M_\odot$ yr$^{-1}$.
Harboring a luminous AGN, the ionizing photons from the nucleus can affect the presence and emission line strengths of PAH particles in F11119+3257. Thus the star formation rate cannot be precisely quantified by measuring PAH emissions. Since the FIR luminosity of F11119+3257 can be derived, we can also evaluate SFR based on the correlations of these two quantities. By checking the IRAS and the SCUBA photometries of F11119+3257 (Table-1), we obtain the FIR luminosity (30-850$\mu$m) of cold dust emission in F11119+3257 $L_{\rm FIR} = 2.1\pm0.3\times10^{46} \rm erg\ s^{-1}$. The contribution of the star-burst activity to the FIR luminosity is estimated to be $\sim 20\%$ based on the restframe $f_{15}/f_{30}$ ratio (ratio between continuum fluxes at 15$\micron$ and 30$\micron$, Veilleux et al. 2013). By applying the correlation of $SFR$($M_\odot$, yr$^{-1}$)=4.5$\pm1.4\times$10$^{−44}$ $L_{\rm FIR}(\rm erg \ s^{-1})$ (Kennicutt 1998), the $SFR$ estimated from the FIR dust emission is $SFR_{\rm FIR} = \x{190\pm90}$ $M_\odot$ yr$^{-1}$. Since the $SFR$ value is more reliably determined based on the FIR luminosity rather than on the PAH emission, it will be applied in the follow-up discussion.

Similar to the optical emission lines, it is reported that the MIR \neiii and \nev emission lines of F11119+3257 are also significantly blueshifted (Spoon et al. 2009). As described in \S3.2.3, we apply the same method in retrieving the \neiii$\lambda$15.56\micron~ and \nev$\lambda$14.32\micron~ OELs (Figure 7.2, 7.3). The MIR \neiii and \nev OELs show a similar two-component profiles to their optical counterparts, their fluxes are also calculated and listed in Table 3.


\begin{figure}[H]
  \begin{center}
      \includegraphics[width=0.87\textwidth]{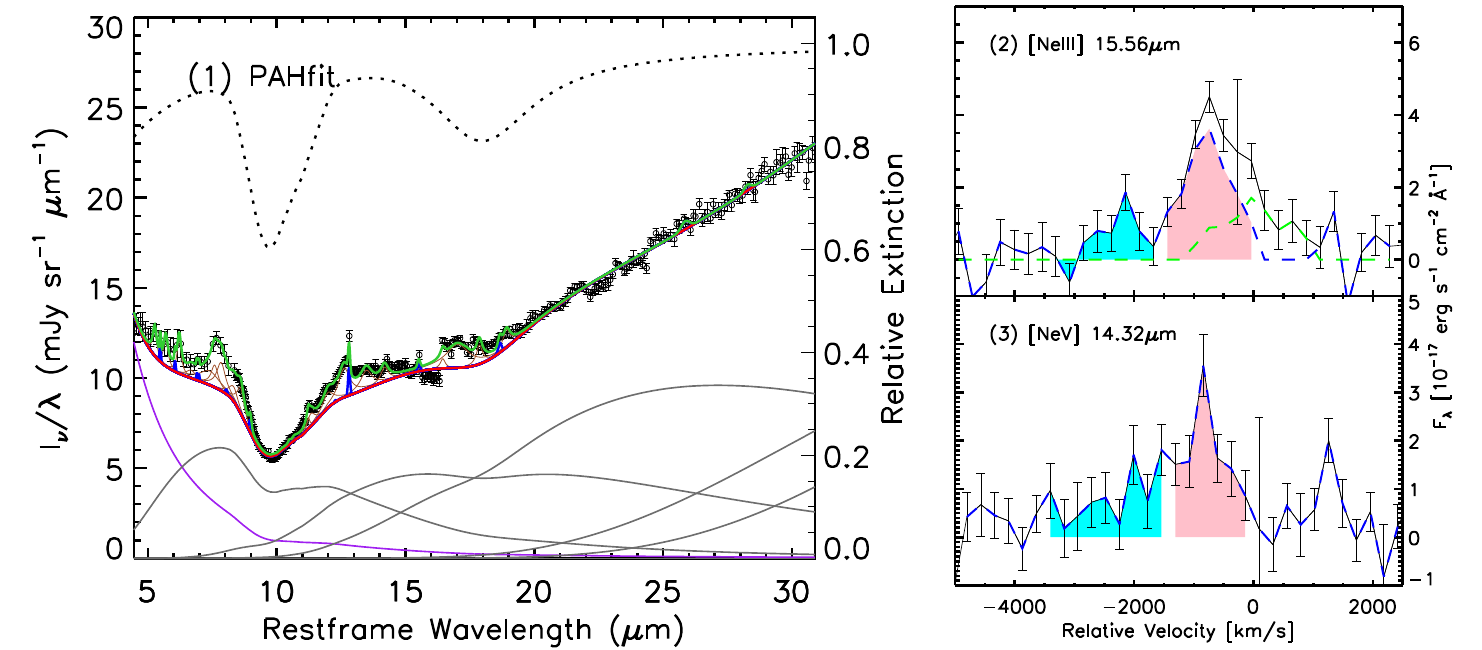}
       \caption{
       Panel (1): PAHfit model of the Spitzer IRS spectrum of F11119+3257. The black 
       circles with error bars are rebinned Spitzer IRS data of F11119+3257.
       Best-fit model is shown with green solid line, combined thermal dust continuum 
       (grey solid lines) and stellar component (purple solid line) is in red solid lines. 
       PAH emissions and other ionic/atomic/molecular lines are presented with brown and blue solid
       lines respectively. All components are reddened with the fully mixed extinction curve
       (dotted black line, with an axis at the right). After subtracting local continuum, the 
       emission line profiles (black solid lines) of \neiii $\lambda$15.56\micron~ and
        \nev$\lambda$14.32\micron~ are displayed in panel (2) and panel (3) respectively. 
       The ``NEL''(green dashed lines), the fast OEL (cyan shaded regions), and the slow OEL
       components (pink shaded regions) are marked in the same way as in Figure 5.
      }
  \end{center}
\end{figure}

\section{DISCUSSION}
\subsection{Properties of the Quasar}
Similar to some previous reports, the SED modeling in \S3.1 reveals that F11119+3257 is heavily obscured in the optical and NIR. Thus the intrinsic optical-NIR fluxes are much stronger than we observed and dereddening needs to be carried out before estimating luminosity and ionizing fluxes.
The BH mass of F11119+3257 has been evaluated in some previous sample studies. One estimate is $10^{7.2} M_\odot$ by Kawakatu et al. (2007), which is based on directly observed monochromatic luminosity at 5100\AA~ and the width of the BELs. The other estimate is $10^{9.44} M_\odot$ by Veilleux et al. (2009), which is based on the $H$ band luminosity. In sample studies, it is a reasonable assumption that the accretion rate is similar among AGNs when BH mass is estimated according to luminosity. However for special AGNs like NLS1s, BH masses estimated with this method tend to be too larger as Eddington ratio of NLS1s are intrinsically higher than that of typical AGNs. Thus it is more reliable to estimate BH mass of F11119+3257 based on scaling relation between the BH and the BELR. Such an empirical correlations are established mostly based on extinction-free AGNs. However, the extinction toward F11119+3257 is likely to be among the strongest of the ``type 1'' AGNs studied in Kawakatu et al. (2007). Unlike sample studies which did not focus mainly on individual objects, we think it is necessary to re-estimate the BH mass after extinction correction in this paper.
Applying the relation of the BH mass with the \halpha line
width and the extinction-free luminosity (Greene and Ho 2005), we estimate that $M_{\rm BH}=1.0^{+0.4}_{-0.2}\times10^8 M_\odot$. Considering the quasar luminosity of $L_{\rm bol} = 10^{12.61} L_\odot \approx 1.6\times10^{46}$ erg s$^{-1}$ (Veilleux et al. 2013), an Eddington ratio of $\frac{L_{\rm bol}}{L_{\rm Edd}} \approx 1.0$ is found, which is typical among NLS1s.
If the MF87 (Mathews \& Ferland 1987) SED of active galaxies is assumed, the incident rate of ionizing photons is $Q_{\rm H} \approx 1.6 \times 10^{56}\rm\ s^{-1}$.

\subsection{Physical Condition of Outflow Material}
Since all the absorption lines lie in a similar velocity range, the blueshifted \nai $\lambda\lambda$ 5891,5897, \ion{He}{1}* $\lambda$ 10830  and \caii $\lambda$3934 mini-BALs are very probably associated with the OH outflow.
Photo-ionization properties of the outflow gas can be constrained with the measured ionic (\caii $\lambda$3934), atomic (\nai $\lambda\lambda$ 5891,5897 and the metastable \hei$\lambda$ 10830) column densities.
An attempt to investigate phyiscal properties of the outflow material is thus conducted with \textit{CLOUDY} (c13.03; Ferland 1998) photo-ionization simulations. 
Given that the line profiles of the broad absorptions are rather complex, the actual gas cloud is likely to be multi-phased. 
However, constrained by the insufficient data quality and limited number of lines being detected, only 1-zone photo-ionization simulation is conducted. The general physical property of the outflow gas could be estimated if simulations converge.

The MF87 AGN SED is assumed as the incident radiation field for the outflowing material in \textit{CLOUDY} simulations.
Since atomic sodium is the most robust column density measurements available, we set up the stopping criteria (thickness of the cloud) for \textit{CLOUDY} to be $N_{\rm Na^0} = 10^{13.7} \rm cm^{-2}$.
With the detection of molecular absorption, the incident radiation field cannot be too strong, or molecules would be destroyed. And a moderate number density of atoms is required to allow molecules to grow.
2-dimensional grids of \textit{CLOUDY} simulations with ionization parameter $U$ ranging from 10$^{-4}$ to 1 and hydrogen number density $n_{\rm H}$ ranging from 1 cm$^{-3}$ to 10$^{10}$ cm$^{-3}$ are carried out. 
Four discrete metallicity values are assumed, i.e. 0.3 $Z_\odot$, 1.0 $Z_\odot$, 3.2 $Z_\odot$, and 10.0 $Z_\odot$, where $Z_\odot$ stands for solar abundance.
Since we found heavy extinction in F11119+3257, dust depletion should be taken into account when dealing with gas-phase absorption.
We assume a depletion pattern of the Galactic warm cloud (Welty et al, 1999), which happens to approximate the depletion pattern observed in several quasar proximate dusty absorbers (Pan et al. 2017; Ma et al. 2017).
The corresponding depletion of Sodium and Calcium are -0.4 $dex$ and -2.0 $dex$ respectively.
Based on the incident radiation ($Q_{\rm H} \approx$ 1.6 $\times 10^{56}\rm\ s^{-1}$) and the definition of ionization parameter $ U = \frac{Q_{\rm H}}{4\pi R^2n_{\rm H}c}$,
 the distance $R$ of the absorber can be figured when $U$ and $n_H$ are specified (thin dotted lines in Figure 8-9).
The matched region between model prediction and measured column densities are shown in Figure 8-9. 
The column density measurement $N_{\rm Ca^{+}} < 10^{14.0} \rm~cm^{-2}$ (thick dashed line in Figure 8-9) puts a constrain of $U \lesssim 10^{-1}$, if a common ISM number density of $n_{\rm H} < 10^6 \rm~cm^{-3}$ is assumed. And the lower limit of log $N_{\rm He^0(2^3S)}$ requires a distance of $R~<~$10 kpc and a number density of $n_{\rm H} \gtrsim 10^2 \rm~cm^{-3}$ in the parameter space investigated.
In addition, if gas-to-dust ratio of the dust-rich MW (Bohlin et al.
1978) or the relatively dust-deficient SMC (Martin et al. 1989) are assumed, the Hydrogen column density corresponding to the measured extinction \ebv = 1.18 is $N_{\rm H}\rm(SMC) = 10^{22.7}~\rm cm^{-2}$ and $N_{\rm H}\rm(MW) = 10^{21.8}~\rm cm^{-2}$ respectively (thich dash-dotted lines in Figure 8-9).
For typical ISM with $n_{\rm H} < 10^6 \rm~cm^{-3}$, the Hydrogen column density of $N_{\rm H}\rm(SMC) = 10^{22.7}~\rm cm^{-2}$ do not overlap with the column density measurements of other species in the parameter space. Thus the MW gas-to-dust ratio is preferred, indicating the outflow gas is relatively dust-rich and has a relatively high metallicity. On the other hand, $N_{\rm H}\rm(MW) = 10^{21.8}~\rm cm^{-2}$ only overlap with other elemental column densities in parameter space with metallicities of $Z\in [0.3,3.2] Z_\odot$. Briefly speaking, these column density measurements suggests that the outflow material posits at a distance of kilo-parsec scale toward the central engine, similar to previous estimates based on the OH mini-BAL (Tombesi et al. 2015).

On the other hand, the blueshifted emission lines are also detected in a similar velocity range (Table~2).
Similar to the outflow in SDSS J1634+2049 reported by Liu et al. (2016), such a resemblance in speed suggests that these blueshifted emission lines probably originate from the absorption line outflow.
We therefore compare the predicted emission line ratios of the \textit{CLOUDY} simulations with the observed fluxes of the outflow emission lines.
There appear to be two distinct OEL component, the fast OEL and the slow OEL (\S3.2.3). Their line ratios are quite different, and may rise from different region of the outflow gas.
Thus these two OEL systems are compared separately with the mini-BAL system.
The strongest OEL, \oiii$\lambda$5007, is used as a normalization to the other emission lines. The matched regions of these line ratios are marked with colored stripes in Figure 8 (slow OEL) and Figure 9 (fast OEL). And since \neiii$\lambda$3869\AA~is probably contaminated by \hei absorption (\S3.2.3), the relatively weaker \neiii$\lambda$3968\AA~ is investigated in stead in the photo-ionization diagnoses.
We found that for the component with higher fluxes, which more likely represents the bulk of the outflow material (i.e. slow OEL), a solution for the emission line ratios can only be found at the solar abundance ($Z = Z_\odot$; Figure 8.2). 
The resulted photo-ionization parameters are log $U \sim~$-2.0, log $n_{\rm H} \sim~$ 2.7, $R \sim~$3.1 kpc and log $N_{\rm H} \sim 21.6~\rm cm^{-2}$(red diamond in Figure 8).
For the fast OEL component (Figure 8), if solar abundance is adopted, a best-fit model is found with log $U \sim~$-1.6, log $n_{\rm H}~\sim~$ 2.5, $R \sim~$2.3 kpc and log $N_{\rm H} \sim 21.8 \rm~cm^{-2}$(black star in Figure 8-9). The corresponding gas-to-dust ratios are close to MW, consistent with the assumption of solar abundance. After a brief comparison between the best-fit parameters of the fast and the slow OEL component, we find that they are generally consistent with each other. 
In the meantime, model results suggest that the fast OEL component might be closer to the central illuminant, and endures a higher incident radiation, which could shield the material more distant in the outflow gas and allow molecules to grow.

\begin{figure}[H]
  \begin{center}
      \includegraphics[width=\textwidth]{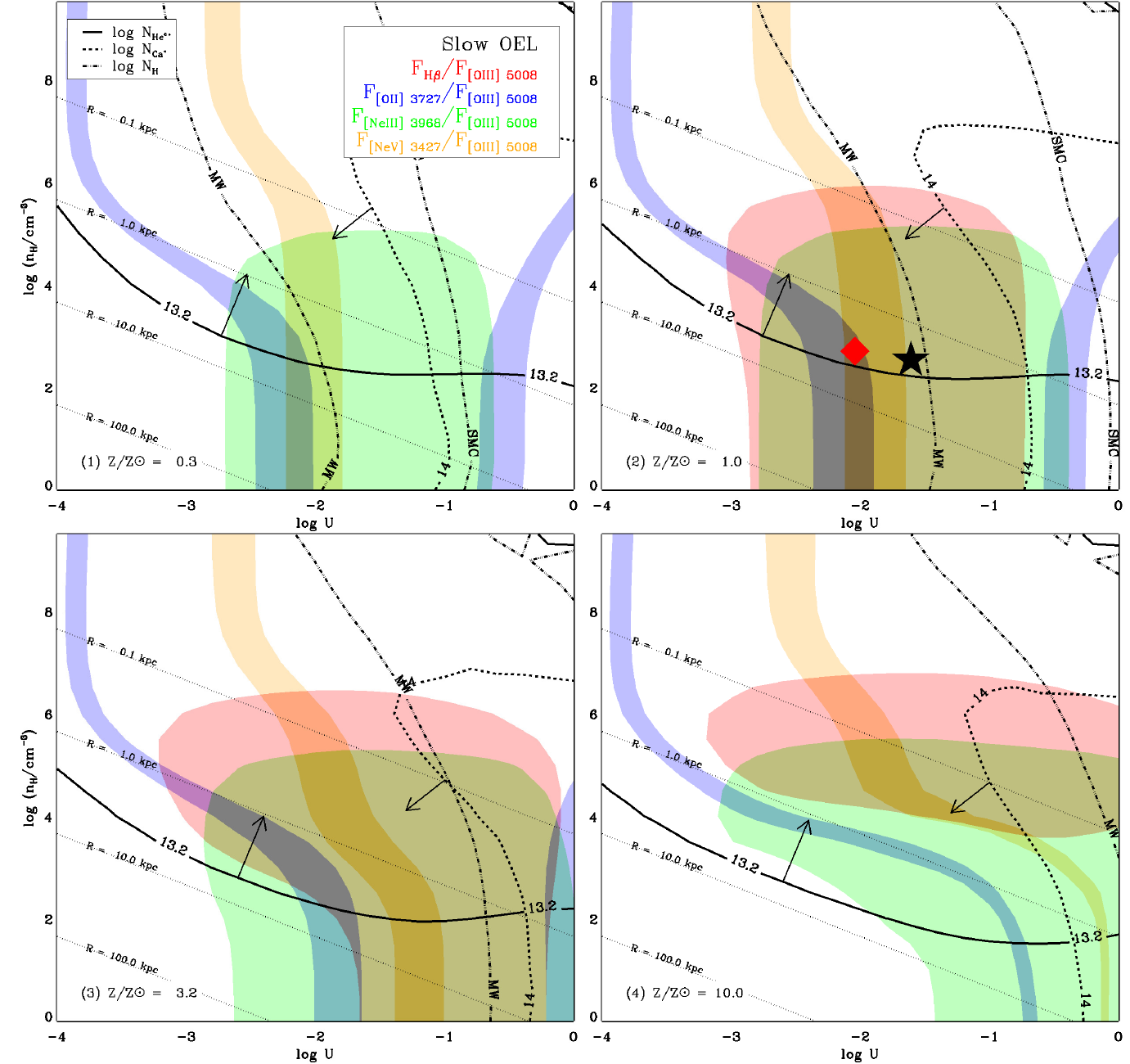}
       \caption{
       \textit{CLOUDY} simulations versus the measured mini-BAL column densities and the
       slow Outflow Emission Line flux ratios are shown. From panel (1) to panel (4),
       the assumed metallicity are 0.3$Z_\odot$, 1$Z_\odot$, 3.2$Z_\odot$, and 10$Z_\odot$
       respectively. Colored stripes correspond to matched emission line ratios between 
       simulations and measurements. The thick dash-dotted lines are Hydrogen column densities
       with assumption of gas-to-dust ratios of MW or SMC. In panel (2) (solar abundance), 
       we mark the best guess of the photo-ionization parameters with a red diamond, which 
       is around log $U \sim$-2.0, log $n_{\rm H}\sim$ 2.7 and $R\sim$3.1 kpc. And the black 
       star is the best-guess for the fast Outflow Emission Lines in Figure-9.
       }
  \end{center}
\end{figure}

\begin{figure}[H]
  \begin{center}
      \includegraphics[width=\textwidth]{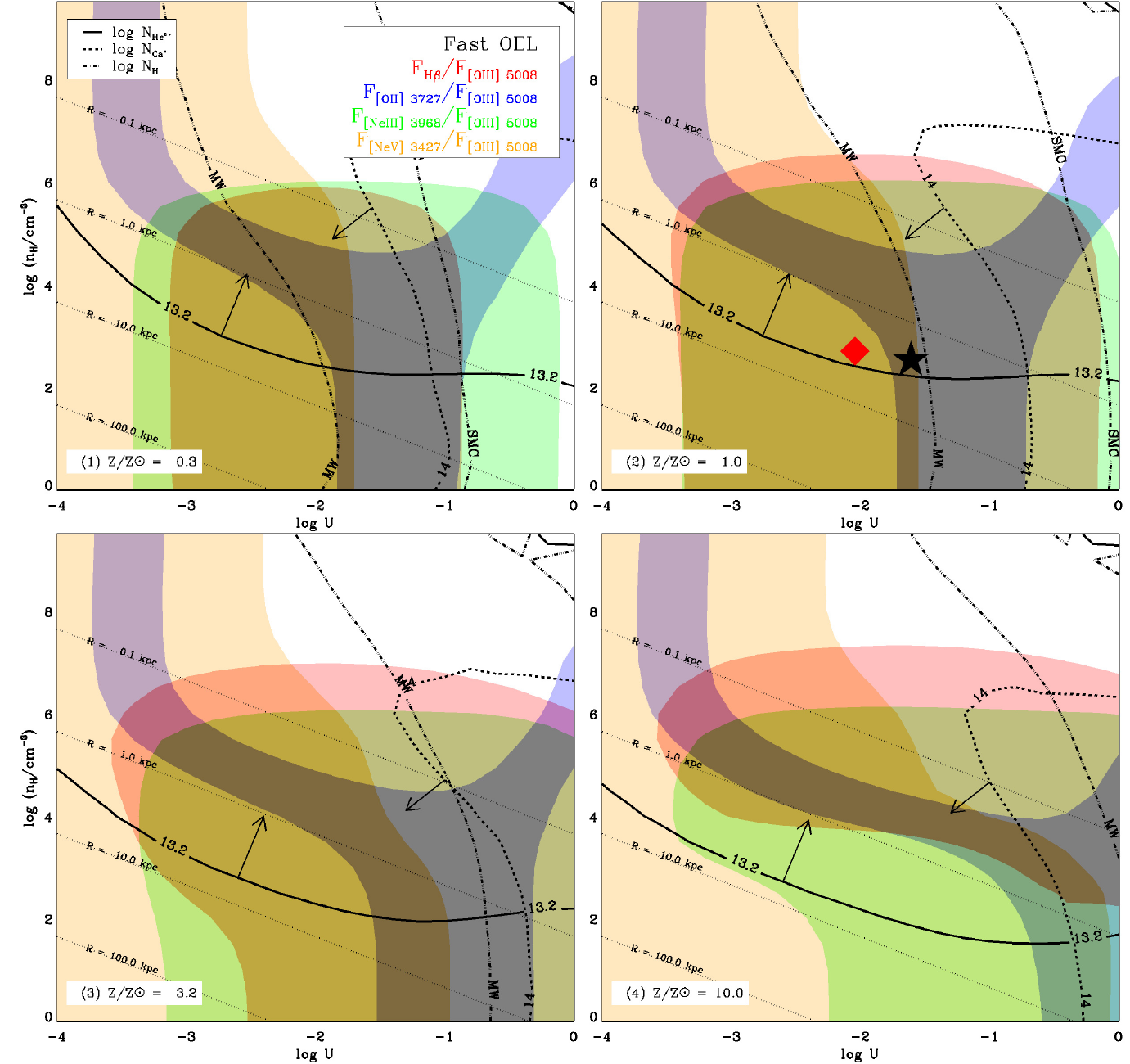}
       \caption{
       Similar to Figure 8, \textit{CLOUDY} simulations versus the measure mini-BAL column densities and 
       the fast Outflow Emission Line flux ratios are shown. In panel (2) (solar abundance), 
       we mark the best guess of photo-ionization parameters with a black star, which 
       is around log $U \sim$-1.6, log $n_{\rm H}\sim$ 2.5 and $R\sim$2.3 kpc. And the red diamond 
       is the best-guess for the slow Outflow Emission Lines in Figure-9.
       }
  \end{center}
\end{figure}

Since the modeled physical parameters of the fast and the slow OEL components are similar, the two sets of results are considered together in estimating the physical condition of the outflow gas. The thickness of outflowing cloud is estimated to be $\Delta R = N_{\rm H}/n_{\rm H} \in [2,12]~$pc, much smaller than the estimated distance $R$.
Thus the bulk of the outflowing material follows roughly a geometry of partially filled thin shell ($\Delta R/R \ll 1$). Following Borguet et al. (2012), the mass-flow rate ($\dot{M}$), momentum flux ($\dot{P}$), and kinetic luminosity ($\dot{E}_k$) can be derived:
$\dot{M} = 4\pi R\Omega \mu m_p N_{\rm H} v,\ $
$\dot{P} = \dot{M} v, \dot{E}_k = \frac{1}{2} \dot{M} v^2$. 
$\Omega$ is the global covering factor of the outflow, and $m_p$ is the proton mass. The flux weighted average velocity of the outflow emission lines are $v\in [-1449,-810]$ \kms~ for different lines (Table 3).
According to $CLOUDY$ simulations, distance and column density are in ranges of
$R \sim$  2.3-3.1 kpc and $N_{\rm H} \sim 10^{\rm 21.6-21.8} \rm cm^{-2}$. Solar abundance is favoured, which corresponds to a mean atomic mass per proton of $\mu = 1.4$. As demonstrated in Liu et al. (2016), strong extinction can lead to large uncertainty in calculating global covering fraction. In the case of F11119+3257, ratio between observed $EW(\rm \oiii)$ and $CLOUDY$ predictions suggests $0.1\% \lesssim \Omega \lesssim 10\%$. The uncertainty is too large for a reasonable estimation of the energetics of the outflow.
Therefore, the relatively more reliable covering factor estimation based on Herschel OH spectrum $C_f(\rm OH) = 0.20\pm0.05$ is adopted (Tombesi et al. 2015). The kinematics of the outflow material is then: $\dot{M} \sim $230-730$ M_\odot\rm yr^{-1}$, the momentum flux is $\dot{P}\sim~$2.2-12.7 $L_{\rm AGN}/c$ and the kinetic luminosity is $\dot{E}_k \sim 10^{43.6-44.8} \sim \x{0.003-0.030}  L_{\rm AGN}.$ 
The estimated distance of the outflow is consistent with recent results based on spatially resolved observations of the CO outflow emission line (at a scale of $\sim 7 ~$kpc; Veilleux et al. 2017).

\subsection{Star Formation in the Host Galaxy}
During the analysis of the emission lines, the narrow Paschen emission (\S3.2.2) discovered in only \paalpha caught our interest. Such narrow emission lines usually come from the NELR of the quasar or from the H II regions around newly formed O stars that traces star formation activity in the host galaxy. 
The Paschen emission is of only half the width of the other NELs, which makes it unlikely originated from the NELR. On the other hand, if we apply the heavy extinction $A_H > 2.1$ (corresponding to \ebv $~>~3.7$) of the Paschen emissions to the NELs we observe, the extinction-free luminosity of the \oii$\lambda$3727 is then $L_{\rm \oii\lambda3727} ~>~5.8\times10^{47}\rm erg s^{-1}$, even larger than the bolometric luminosity of the quasar. Thus the Paschen emissions are unlikely NELs of the quasar, which makes star formation region the most plausible origin. Assuming a case-B line ratio for the anomalous Hydrogen emission lines, the luminosity of \halpha associated with the narrow Paschen component after extinction correction is $L_{\rm \halpha peak}~>~1.7\times^{43}\rm erg s^{-1}$. An $SFR$ of $SFR_{\rm \paalpha peak}~>~130~M_\odot$ yr$^{-1}$ is obtained according to the relation $SFR = L_{\rm \halpha}/1.26\times10^{41}\rm~erg s^{-1}$ (Kennicutt 1998). Since extinction is tightly correlated with $SFR$ as studied in star-forming galaxies (Xiao et al.2012), the heavy extinction to the narrow Paschen emission is expected as the star formation rate is high.
The lower limit of $SFR$ based on the anomalous Paschen emission is \x{of a similar order of our previous estimates based on FIR emission ($SFR_{\rm FIR} = $190$\pm$90$~M_\odot$ yr$^{-1}$)}, and the extinction to the star formation region is also present in the Spitzer spectrum as the prominent silicate absorption.
Detection of the narrow \paalpha emission indicates that spectroscopy of the IR Hydrogen emission lines can be powerful probes to the star-burst regions that are usually heavily obscured. With the launch of JWST, the narrow Brackett (1.46-4.05\micron) and Pfund (2.28-7.46\micron) emissions from H II region can be observed in detail, which will help to precisely measure the obscuration and spatial distribution of the star formation region.

In the previous section, we show that large amount of kinetic energy is injected into the host galaxy in the form of powerful outflows. Such strong galactic-scale feedback in F11119+3257 can be responsible for the violent star formation activity, or regulate star formation processes.
On the other hand, stellar feedback at high $SFR$ can also play a significant role in forming outflows. The simulation in Biernacki \& Teyssier (2018) shows that, in a dark matter hole of the mass around 2$\times 10^{12} M_\odot$, continuous $SFR$ at $\sim 100~M_\odot \rm yr^{-1}$ along can launch massive ($\sim 20~M_\odot \rm yr^{-1}$) molecular outflow at a scale of $\sim$ 20 kpc. At $\sim $\x{190}$~M_\odot \rm yr^{-1}$, the more violent $SFR$ in F11119+3257 are capable of contributing significantly to the massive molecular outflow been observed. According to some previous findings (Dav{\'e} et al. 2011, etc), however, the fast wind speed of the outflow ($\sim1000 $\kms) in F11119+3257 can only be caused by AGN feedback.
Given the power and scale of the outflow, it must interact with the violent star formation been observed. Detailed processes of the interactions need follow-up spatially-resolved observations and simulations to reveal.

\subsection{Implications of MIR Emission Lines}

\begin{figure}[H]
  \begin{center}
      \includegraphics[width=0.5\textwidth]{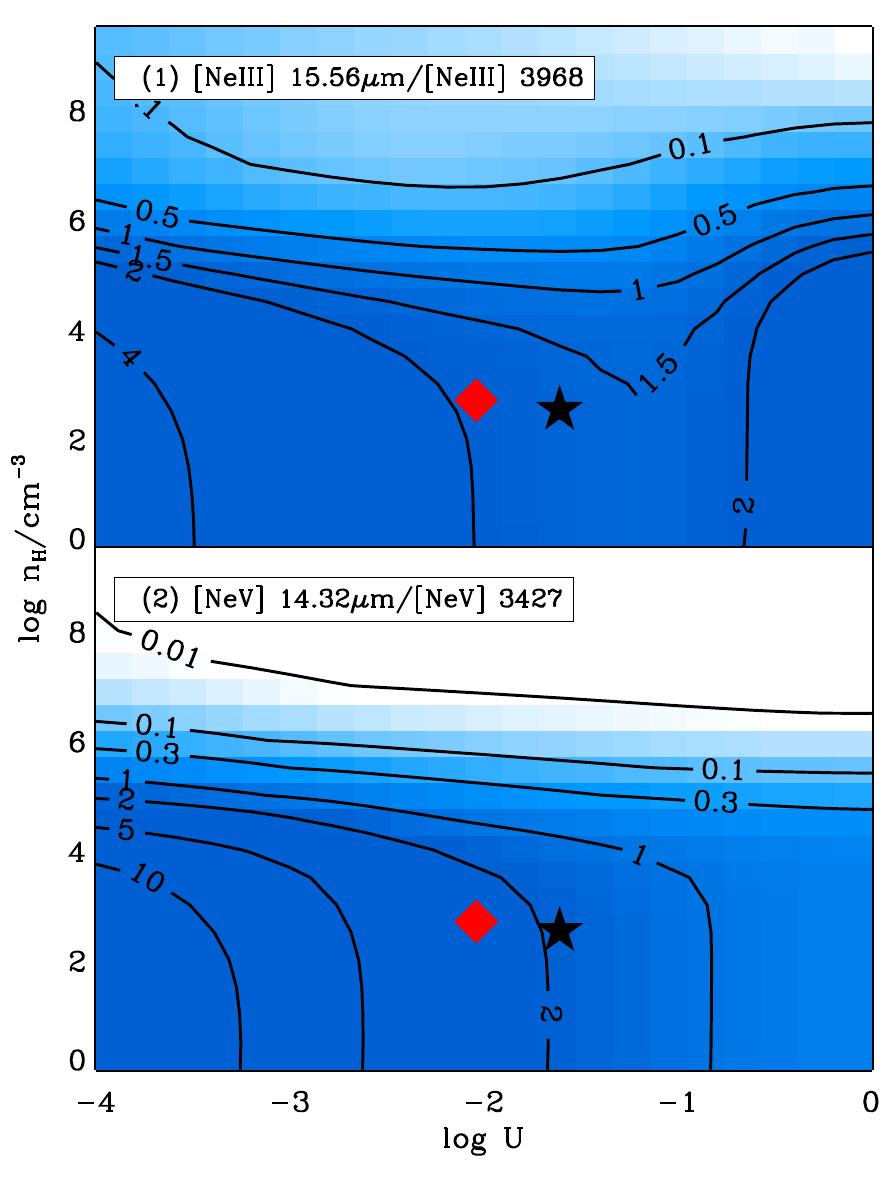}
       \caption{
       The emission line ratios of \neiii$\lambda$15.56\micron/\neiii$\lambda$3968 (upper panel) and
       \nev$\lambda$14.3\micron/\nev$\lambda$3427 (lower panel) predicted by the \textit{CLOUDY} 
       simulations with solar abundance
       are shown. The predicted line ratios for the OELs (around the red squares and black stars) are
       \neiii$\lambda$15.56\micron/\neiii$\lambda$3968(\textit{CLOUDY})$\approx$1.8 and
       \nev$\lambda$14.32\micron/\nev$\lambda$3427(\textit{CLOUDY})$\approx$2.0.
       }
  \end{center}
\end{figure}

As shown in section 3.4, prominent blueshifted MIR Neon lines are detected in F11119+3257. It is an interesting fact that the line profile of the MIR Neon emission lines are similar to optical Neon lines, suggesting that both of them should rise from the outflow gas. A comparison study of the optical and the MIR Neon lines can probably implicate physical properties of the emission line region. Given the large difference in wavelength between the optical ($\sim$0.4\micron) and the MIR ($\sim$15\micron) Neon lines, potential extinction to the outflow gas can be investigated since extinction at the optical bands are much stronger than in the MIR. Heavy extinction is found for the both the continuum and the BELs of F11119+3257, the OELs can also be obscured. The observed OEL fluxes ratios are \neiii$\lambda$15.56\micron/\neiii$\lambda$3968(observed)$\approx$13.7 and \nev$\lambda$14.32\micron/\nev$\lambda$3427(observed)$\approx$16.8. However, the predicted emission line ratios for the OELs \x{by} \textit{CLOUDY} simulations are only around \neiii$\lambda$15.56\micron/\neiii$\lambda$3968(\textit{CLOUDY})$\approx$1.8 and \nev$\lambda$14.32\micron/\nev$\lambda$3427(\textit{CLOUDY})$\approx$2.0 (Figure 10). The relatively lower fluxes of the blueshifted Neon lines at the optical bands are very likely caused by dust extinction given the association of the outflow with the obscuring dust. If the cubic extinction curve for the continuum is applied, an extinction of \ebv$^{\rm OELR}\approx$~0.5 for the outflow can be derived. This value is lower than the extinction to the BELR, consistent with the changing profiles of the Hydrogen emission lines been observed (OELs become stronger at shorter wavelengths as compared to BELs; \S3.2.2).

\begin{figure}[H]
  \begin{center}
      \includegraphics[width=\textwidth]{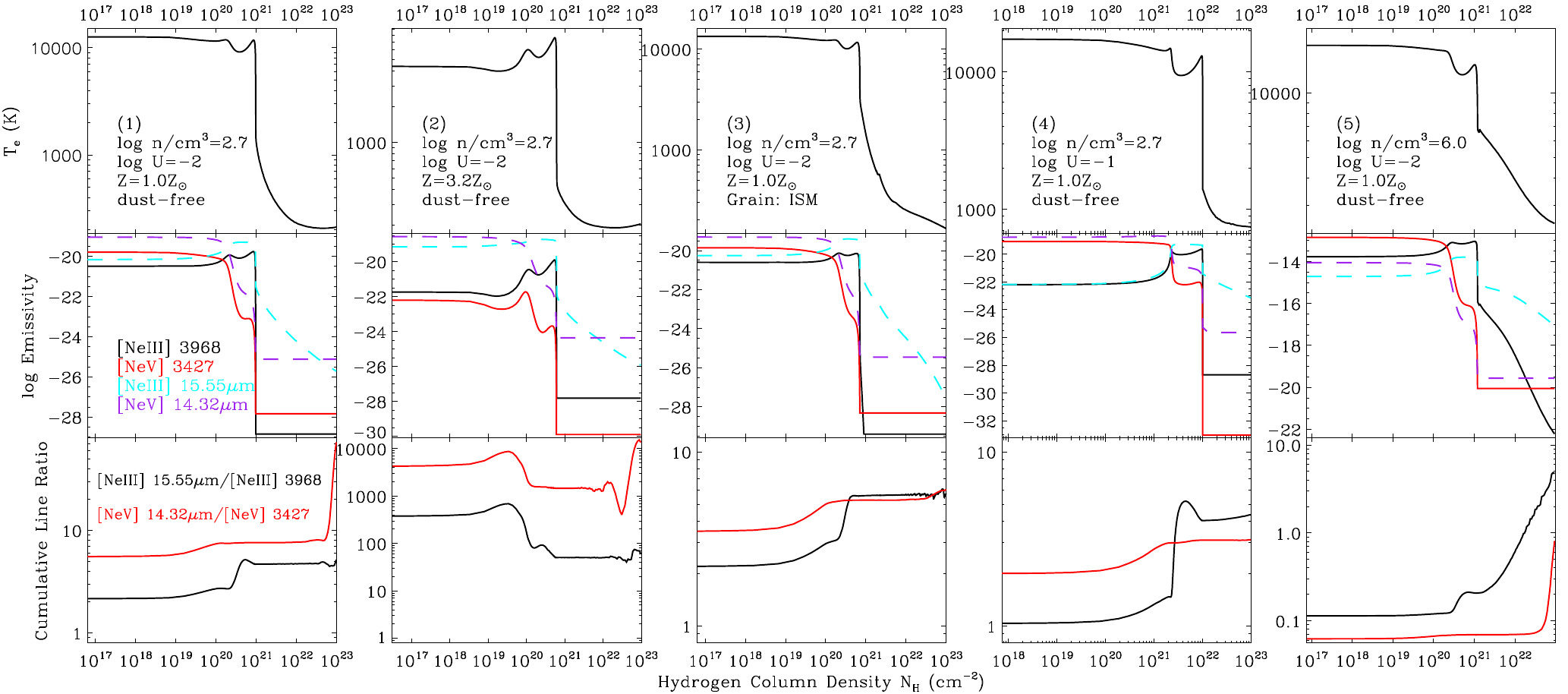}
       \caption{
       Emissivity structure of Neon lines in clouds illuminated by AGN predicted by \textit{CLOUDY} simulations.
       The structure of electron temperature (T$_e$) is in the upper panels, line emissivities are in the 
       middle panels and cumulative line ratios are in the lower panels. The differences between panel (1) 
       and other panels are: (2) metallicity, (3) presence of dust, (4), ionization parameter, (5) number 
       density.
       }
  \end{center}
\end{figure}

On the other hand, the MIR-to-optical Neon line intensity ratios can also be affected by factors other than extinction as upper levels of these transitions are different. \textit{CLOUDY} simulations of emissivity structures of these Neon lines in clouds illuminated by AGN reveal some clue. Thick gas slabs (log N$_{\rm H}/\rm cm^{-2} = 23$) with various physical conditions are assumed in \textit{CLOUDY} modelings, and results are shown in Figure 11.1-5 respectively. The structures of temperature (T$_e$), line emissivities and cumulative line ratios are in the upper, middle and lower panels respectively. First of all, as number density rises from n$_{\rm H} = 10^{2.7}\rm cm ^{-3}$ (Figure 11.1) to about n$_{\rm H} = 10^{6}\rm cm ^{-3}$ (around the critical density of these Neon lines; Figure 11.5), the line emissivities rise dramatically, and the cumulative line ratios show an entirely different pattern. Beside number density, it is clear that the emissivity and line ratios are mainly determined by temperature of the gas, which is affected by facts like ionization parameter $U$ (Figure 11.1 v.s. Figure 11.4), Metallicity $Z$ (Figure 11.1 v.s. Figure 11.2), and the presence of dust (Figure 11.1 v.s. Figure 11.3). In the case of the outflow gas in F11119+3257, the observed line ratios of \neiii$\lambda$15.56\micron/\neiii$\lambda$3968(observed)$\approx$13.7 and \nev$\lambda$14.32\micron/\nev$\lambda$3427(observed)$\approx$16.8 are hard to be explained as caused by the change of the physical conditions as shown in the figures. The variation of neither $n_{\rm H}$, $U$, or dust content can boost the MIR/optical Neon ratios to be over 10. A higher metallicity (Figure 11.2) though, can result in a \neiii$\lambda$15.56\micron/\neiii$\lambda$3968~ ratio similar to our observation, but the predicted \nev$\lambda$14.32\micron/\nev$\lambda$3427~ intensity ratio becomes much too large. Therefore the case of F11119+3257 implicates that comparison study of MIR and optical Neon lines can provide additional diagnoses to the physical conditions (e.g. extinction) of the emission line regions of galaxies.

\section{SUMMARY AND PROSPECT}
In this paper, we carry out a comprehensive study of the continuum, emission lines, and absorption liness of the fast molecular outflow candidate, a ULIRG + partially-obscured NLS1 IRAS F11119+3257. A systemic redshift of $z = 0.18966\pm0.00006$ is determined by modeling low-ionizaiton narrow emission lines. And significant dust extinction at \ebv$\sim$ 1.18 is found for both the continuum and the BELs. The dereddened optical spectrum reveals a powerful AGN with a black hole mass of $M_{\rm BH} = 1.0^{+0.4}_{-0.2}\times10^8~M_\odot$, which accretes at \x{around the Eddington rate} of $\frac{L_{\rm bol}}{L_{\rm Edd}} \x{\approx 1}$. Broad absorption lines from ions (\caii$\lambda$3934), neutral atoms (\nai$\lambda\lambda$5891,5897  and the metastable \hei$\lambda$10830 ) are detected at the velocities ($\sim -1000$ \kms) similar to that of the molecular OH 119\micron\ mini-BAL. Photo-ionization diagnosis of the ionic, atomic, and molecular lines found that the mini-BAL outflows are at kilo-parsec scale, consistent with previous reports.
At similar speeds to the mini-BAL, a blueshifted emission line system is identified in several ions (\oiii$\lambda\lambda$4959,5007, \oii $\lambda$ 3727, \neiii$\lambda$3869, and \nev$\lambda$3427) and atomic Hydrogen (\halpha, \hbeta and \hgamma). The similarity in kinetic properties between the blueshifted emission and absorption lines suggests that they could be associated.
The blueshifted emission lines are broad, and are divided into two components: the slow OEL component ((-1500,0] \kms) and the relatively high-ionization fast OEL component ([-3500,-1500] \kms).
After inspecting the predicted emission line ratios of the mini-BAL outflows, we find that it will match the line ratios of the fast and slow OEL components if solar abundance and a dust-to-gas ratio similar to MW are assumed. And the physical conditions of the fast and slow OEL components \x{are} similar even they differ largely in their line ratios. These results indicate that the OELs are very likely originated from the mini-BAL outflows.
A joint analysis based on these ionic/atomic/molecular mini-BALs and OELs reveals the position ($R \sim \x{3}~$kpc from the black hole) and physical properties ($n_H \sim 10^{2.6}\rm~cm^{-3}$ and $N_H\sim 10^{21.7} \rm~cm^{-2}$) of the galactic scale outflow.
Further more, the mass-flow rate ($\dot{M}\sim \x{230-730}~M_\odot \rm yr^{-1}$), the momentum flux ($\dot{P} \sim \x{2.2-12.7} ~L_{\rm AGN}/c$) and the kinetic luminosity ($\dot{E}_k \sim 10^{\x{43.6-44.8}} \sim \x{0.003-0.030}~L_{\rm AGN}$) are obtained. In addition, the presence of the blueshifted \neiii and \nev emission lines in the MIR provides chances to investigate additional properties of the outflow gas, and an extinction of \ebv $\sim$ 0.5 is inferred for the outflow region.
Besides, we found that the Paschen emission line in F11119+3257 is unique, in which a very narrow ($FWHM_{\rm \paalpha peak} = 260\pm$20~\kms~ $~<~ FWHM_{\rm NEL}/2$) emission line component is found only in \paalpha. By checking this Paschen emission line component in \paalpha and \pabeta, a heavy extinction at $A_H > 2.1$ (corresponding to \ebv $~>~3.7$) is found. The most plausible explanation to this anomalous Paschen emission is that it comes from heavily obscured star formation, and a lower limit of $SFR_{\rm \paalpha peak}~>~130M_\odot$ yr$^{-1}$ is obtained. The result is \x{of a similar order of the estimates inferred from FIR emissions ($SFR_{\rm FIR} = $190$\pm$90$~M_\odot$ yr$^{-1}$)}. This shows that Hydrogen emission lines in the IR (Paschen series, Bracket series, etc.) can be powerful probes to the star formation regions even if they are heavily obscured. 

With such a massive modelcular outflow and high $SFR$, F11119+3257 is among the best laboratory to study both the formation of galactic-scale outflows and the outflow-starformation interactions.
At a physical scale of $\sim3$kpc ($\sim1\arcsec$ for $z =~$0.18966), the emission lines from both the galactic-scale outflow  and the star formation region can be spatially resolved by the next-generation telescopes like JWST. The spatially and spectrally resolved observations will enable a clear view of the outflow and the star formation processes, and their interactions.

\acknowledgments
We are thankful to the helpful comments about starformation rate estimation and the implications of MIR Neon emission lines from an anonymous referee on this paper.
This work is supported by the National Natural Science Foundation of China (NSFC-11573024, 11903029, 11473025, 11421303) and the National Basic Research Program of China (2013CB834905). WL acknowledges supports from the Natural Science Foundation of China grant (NSFC-11703079) and the "Light of West China" Program of Chinese Academy of Sciences (CAS). PJ is supported by the National Natural Science Foundation of China (NSFC-11973037) and HW is supported by the Natural Science Foundation of China (NSFC-11890693). We acknowledge the use of the Hale 200-inch Telescope at Palomar Observatory through the Telescope Access Program (TAP), as well as the archival data from the SDSS, 2MASS, Catalina, WISE surveys and  Spitzer space telescope. TAP is funded by the Strategic Priority Research Program, the Emergence of Cosmological Structures (XDB09000000), National Astronomical Observatories, Chinese Academy of Sciences, and the Special Fund for Astronomy from the Ministry of Finance. Observations obtained with the Hale Telescope at Palomar Observatory were obtained as part of an agreement between the National Astronomical Observatories, Chinese Academy of Sciences, and the California Institute of Technology. Funding for SDSS-III has been provided by the Alfred P. Sloan Foundation, the Participating Institutions, the National Science Foundation, and the U.S. Department of Energy Office of Science. The SDSS-III website is http://www.sdss3.org/

\clearpage

\begin{deluxetable}{rrccc}
\tabletypesize{\normalsize} \tablecaption{Photometric and Spectroscopic Observations
\label{tbl-1}} \tablewidth{0pt} \tablehead{
\colhead{Band} & \colhead{Magnitude$\vert$Flux} & \colhead{Survey/Telescope} & \colhead{Obs. Date} &\colhead{Ref.}\\
\colhead{ } & \colhead{\ \ \ \ \ \ \ \ [mag$\vert$Jy]} & & \colhead{(UT)} & \colhead{ } }
\startdata
 &  & Photometries \\
\hline
  u(PSF)  & $21.25\pm0.10$ mag  & SDSS  &  2004 Apr 16 & 1,2      \\ 
  g(PSF)  & $19.02\pm0.02$ mag  & SDSS  &  2004 Apr 16 & 1,2\\
  r(PSF)  & $17.21\pm0.01$ mag  & SDSS  &  2004 Apr 16 & 1,2\\
  i(PSF)  & $16.11\pm0.02$ mag  & SDSS  &  2004 Apr 16 & 1,2 \\
  z(PSF)  & $15.84\pm0.02$ mag  & SDSS  &  2004 Apr 16 & 1,2 \\ 
 $J$      & $14.08\pm0.03$ mag & 2MASS & 2000 Jan 02 & 3      \\
 $H$      & $12.91\pm0.03$ mag & 2MASS & 2000 Jan 02 & 3   \\
 $K_{s}$  & $11.60\pm0.02$ mag & 2MASS & 2000 Jan 02 & 3 \\  
  $W1$      & $9.92\pm0.01$ mag & $WISE$  & 2010 May 23 & 4\\
 $W2$     & $8.84\pm0.01$ mag & $WISE$  & 2010 May 23 & 4\\
 $W3$     & $6.17\pm0.01$ mag & $WISE$  & 2010 May 23 & 4\\
 $W4$    & $3.63\pm0.01$ mag  & $WISE$  & 2010 May 23 & 4\\  
 25$\mu$m & $0.348\pm0.022$ Jy & IRAS & 1983 & 5\\
 60$\mu$m & $1.59\pm0.18$ Jy & IRAS & 1983 & 5\\
 100$\mu$m & $1.52\pm0.17$ Jy & IRAS & 1983 & 5\\ 
 850$\mu$m & $5.6\pm1.9$mJy & SCUBA & 2003 Jan 18 & 6\\
1.4GHz    & $106.07\pm0.12$ mJy & $FIRST$ &  1994 Jun 14 & 7 \\ 
\hline
& & Spectroscopies\\
\hline
3610--10390 [\AA] & \ldots & SDSS/BOSS  & 2013 Mar 16 & 8 \\
9800--24500 [\AA] & \ldots & P200/TSpec & 2013 Feb 24 & This Work\\%
9000--20790 [\AA]& \ldots & VLT/ISAAC &  2013 Mar 20 & 9\\
4.4--38.2[$\mu$m]& \ldots & Spitzer/IRS & 2004-May-11 & 10\\ 
\enddata
\tablerefs{
(1) York et al. 2000; (2) Abazajian et al. 2009; (3) Skrutskie et al. 2006; (4) Wright et al. 2010; (5) Moshir \& et al. 1990; (6)Clements et al. 2010; (7) White et al. 1997; (8) Ahn et al. (2014); (9) ESO program: 088.B-0997(A); (10) Higdon et al. 2006, reduced data: Lebouteiller et al. 2011; 
}
\end{deluxetable}

\begin{deluxetable}{lrrccccccccc}
\tabletypesize{\footnotesize} \tablecolumns{4} \tablewidth{0pt}
\tablecaption{Emission Line Components} \tablehead{
\colhead{Component} &\colhead{$v_0^{\rm (a)}$} &\colhead{$FWHM$} &\colhead{Line Profile} \\
\colhead{}          &\colhead{\kms}  &\colhead{\kms}   &\colhead{}
} \startdata
\feii\ BEL       &   10$\pm$20    &\x{1800$\pm$200 } &  Lorentzian     \\
\feii\ NEL       &  120$\pm$10    & 740$\pm$ 50  &  Gaussian       \\
NEL             &    0$\pm$10    & 570$\pm$ 40  &  Gaussian       \\
\paalpha$^{(b)}$&    0$\pm$10    & 890$\pm$ 50  &  \ldots     \\
\pabeta$^{(b)}$ &   0$\pm$30    &\x{1100$\pm$300}  &  \ldots     \\
\halpha$^{(b)}$ &  -70$\pm$ 10    &1310$\pm$ 40  &  \ldots     \\
\hbeta$^{(b)}$  & \x{-130$\pm$20}    &\x{1450$\pm$130 } &  \ldots     \\
\enddata
\tablecomments{\\(a) Relative velocity in the quasar's rest frame, $z=0.18966$. \\
(b) Centroids and $FWHM$s of \feii multiplets are derived from model parameters, centroids and $FWHM$s of \halpha, \hbeta, \paalpha, \pabeta emission lines are directly measured.}
\end{deluxetable}

\begin{deluxetable}{lrrrcccccccc}
\tabletypesize{\footnotesize} \tablecolumns{4} \tablewidth{0pt}
\tablecaption{Outflow Emission Line Properties} \tablehead{
\colhead{Transition} &\colhead{Flux$_{\rm Fast\ OEL}$} &\colhead{Flux$_{\rm Slow\ OEL}$} &\colhead{Average Velocity $\bar{v}^{(a)}$} \\
\colhead{}          &\colhead{10$^{-17}$ erg s$^{-1}$ cm$^{-2}$}  &\colhead{10$^{-17}$ erg s$^{-1}$ cm$^{-2}$}   &\colhead{\kms}
} \startdata
\oiii$\lambda$5007   & 650$\pm$90 &  1230$\pm$80  &  \x{-1330}\\
\hbeta              & 110$\pm$90 &   110$\pm$60  &  \x{-1640}\\
\oii$\lambda$3727    &  20$\pm$10 &   170$\pm$30  &   \x{-810}\\
\neiii$\lambda$3869  &  65$\pm$40 &    $>$90     &  \x{-1200}\\
\neiii$\lambda$3968  &  50$\pm$30 &    60$\pm$30     &  \x{-1420}\\
\nev$\lambda$3427    &  30$\pm$20 &    70$\pm$20  &  \x{-1260}\\
\neiii$\lambda$15.56\micron &  \x{500$\pm$400} &    \x{1700$\pm$200}     &  \x{-960}\\
\nev$\lambda$14.32\micron &  \x{600$\pm$500} &    \x{1000$\pm$300}     &  \x{-810}\\
\enddata
\tablecomments{\\(a). Flux weighted relative velocity in the quasar's rest-frame. }
\end{deluxetable}

\clearpage

\end{document}